\documentstyle[pre,aps,epsfig, eqsecnum,array,multicol]{revtex}

\def\top#1{\vskip #1\begin{picture}(290,80)(80,500)\thinlines \put(
65,500){\line( 1, 0){255}}\put(320,500){\line( 0, 1){
5}}\end{picture}}
\def\bottom#1{\vskip #1\begin{picture}(290,80)(80,500)\thinlines \put(
330,500){\line( 1, 0){255}}\put(330,500){\line( 0, -1){
5}}\end{picture}}

\begin{document}
\draft

\title{Theory of monolayers with boundaries: Exact results and
Perturbative analysis}

\author{Joseph Rudnick and Kok-Kiong
Loh} \address{Department of Physics, University
of California at Los Angeles, California
90095-1547} 
\date{\today} 
\maketitle

\begin{abstract}
Domains and bubbles in  tilted phases of Langmuir
monolayers contain a class of textures known as boojums.  The
boundaries of such domains and bubbles may display either cusplike
features or indentations.  We derive analytic expressions for the
textures within domains and surrounding bubbles, and for the shapes of
the boundaries of these regions.  The derivation is perturbative in
the deviation of the bounding curve from a circle.  This method is
not expected to be accurate when the boundary suffers large
distortions, but it does provide important clues with regard to the
influence of various energetic terms on the order-parameter texture
and the shape of the domain or bubble bounding curve.  We also look
into the effects of thermal fluctuations, which include a
sample-size-dependent effective line tension.
\end{abstract}

\begin{multicols}{2}

\section{Introduction}
Monolayers of insoluble surfactant molecules confined to the air/water
interface possess complex phase structures \cite{phases}.  In the
``tilted'' phases, the long axes of the surfactant molecules in the
monolayer are uniformly tilted with respect to the normal and 
the molecular tilt azimuth organizes spontaneously on
macroscopic length scales.  The structures adopted by the molecular
tilt azimuth are referred to as {\em textures}.  There is no
long-range order of the tilt azimuth in the liquid expanded ($LE$) and
the gaseous ($G$) phases.  Tilted phases can coexist with the $LE$ and
$G$ (isotropic) phases and form micron-sized domains.  Alternatively,
bubbles of an isotropic phase may appear against a background of a tilted
phase.  Nontrivial textures in the domains, and around the bubbles,
have been observed in the $L_2/LE$ and $L_2/G$ coexistence region,
where the $L_2$ phase is one of the tilted phases.
Boojum textures, similar to those seen in superfluid $^3$He\cite{Mermin} 
and smectic-$I$ (Sm-$I$) droplets in liquid-crystal
films\cite{LanSeth}, have been observed in the interior of $L_2$
domains \cite{RivMeu}. 
An ``inverse boojum,'' which is the texture around the bubble analogous to 
the boojum in the case of the domain, has been identified \cite{FangTeer}.
The domains and bubbles associated with boojums are not circular.
Among the nontrivial domain shapes seen are protrusions on both
bubbles and domains, at times sharp enough to be characterized as
``cusps''\cite{RivMeu,FangTeer} and indentations
in domain boundaries which impart a cardioid appearance to the domain
\cite{cardioid,cigar}.  Such domains and bubbles with unusual textures and
shapes ought to be observable in other tilted phases as well.

The above textures can be understood in terms of continuum elastic
theories of smectic liquid crystals\cite{FiscBru}.  The bulk energy is
controlled by elastic moduli that quantify the energy cost of
bend and splay distortions.  There are also contributions to the
boundary energy, known as the line tension, that depend on the
relative angle between the boundary normal and the tilt azimuth.  In
equilibrium, the texture in a domain or surrounding a bubble, and
the shape of the boundary between condensed and expanded regions,
adjust so as to minimize the total energy of the monolayer.  Domains
with nontrivial textures and shapes represent the compromise arising
from the competition between the bulk energy and the line tension.

Simultaneous determination of the texture and the boundary of the
domain poses a calculational challenge.  Earlier studies
include the exact result discovered by Rudnick and Bruinsma for a
domain with isotropic elastic energy and only the first anisotropic
contribution in the Fourier expansion of the line tension $\sigma(\phi)$
\cite{RudBru}, and the perturbation about the exact result in terms of
coefficient of the second anisotropic term in the
expansion\cite{RudBru}.  Galatola and Fournier have approached the
problem of domains with elasticity and line-tension anisotropies by
searching numerically for the equilibrium domain shapes and positions
of domains in a fixed texture background~\cite{GalaFour}.  Rivi\`{e}re
and Meunier \cite{RivMeu} have attributed their experimental findings
on domain shapes and textures to elasticity anisotropy in the same
manner as in Ref.  \cite{GalaFour}.  In the work of Fang {\em et
al.} \cite{FangTeer}, nontrivial boundary shapes for both domains and
bubbles, as well as the ``inverse boojum'' textures in the $L_2$ phase
surrounding the bubbles, have been reported.  A brief account of the
theoretical understanding of the bubbles has also been presented in
Ref.~\cite{FangTeer}.

In this work, we extend the effort of Rudnick and Bruinsma
\cite{RudBru} to analyze the problem of domains with anisotropic
elastic energy in addition to the line-tension anisotropy.  We also
generalize the approach to the problem of bubbles and provide a
detailed derivation of the results that have been published in
Ref.~\cite{FangTeer}.  Careful analysis reveals that although
protrusions can be expected to form on the boundary of a domain of the
$L_2$ phase, a ``cusp'' in the form of a discontinuity in the slope of
the bounding curve surrounding the domain will not appear in the
parameter regime that is appropriate to the analysis
 that has been carried out~\cite{RudBru,GalaFour,FangTeer}.
The conclusion  in Ref.~\cite{RudBru} that a cusp exists is
due to an approximation\cite{private} that affects the qualitative 
results of the analysis.  The fact that the cusp does not
exist and the condition for the existence of cusps
on the boundary were first pointed out by Galatola and Fournier
\cite{GalaFour}.  A formal
derivation of the conditions for the appearance of a cusp will be
provided in this paper. Perturbative results, which yield the
effects of small anisotropies on the textures and boundaries, are
obtained.  The reliability of the perturbative approach when the
boundary is significantly different from a circle is not obvious.
Nevertheless, one is provided with useful insights with regard to the
influence of various contributions to the energy of the Langmuir
monolayer.  In addition, we examine the effect of thermal
fluctuations.  We are led to a renormalized line tension that depends
on the radius of the boundary\cite{Saleur}.

We have also implemented a numerical program using finite
element methods for evaluation of the equilibrium texture and boundary
simultaneously.  With the use of this approach, we are able to explore
regions of the parameter space that are not accessible to the
perturbative technique.  A brief report on the numerical work has
already appeared \cite{numerics}.  A full description of this method
and a systematic review of the results of its implementation are
deferred to a future article.

The organization of this paper is as follows.  In Sec.
~\ref{sec:approach}, we describe the approach in general.  In
Sec.~\ref{sec:exactsolutions}, we summarize the exact analytic
results.  Section~\ref{sec:textureboundary} displays the perturbative
analysis of the relation between the texture and the boundary.
Sections~\ref{sec:moredomain} and~\ref{sec:bubbles} describe the
analysis for the cases of domains and bubbles that results from
perturbing about the exact solutions.  In Sec.~\ref{sec:thermal},
we analyze the effect of thermal fluctuations.  Concluding remarks are
contained in Sec.~\ref{sec:conclusions}.

\section{The Approach}
\label{sec:approach}
We describe the monolayer by an ordered phase with $XY$-like order parameter 
$\hat{c}(x, y)=\hat{x}\cos\Theta(x, y)+\hat{y}\sin\Theta(x, y)$, a 
two-dimensional unit vector indicating the direction
of the projection onto the substrate of the tilted hydrophobic tail
of the surfactant forming the Langmuir monolayer. The quantity, $\Theta(x,
y)$, is the angle that 
$\hat{c}(x,y)$ makes with the $x$ axis.
When a region $\Omega$ contains an ordered phase which is invariant under
in-plane reflection, the free energy of the system
takes the general form~\cite{FiscBru}
\begin{eqnarray}
H[\Theta(x, y)] = &&\int_\Omega{\cal H}_b\; dA+ \oint_\Gamma \sigma
[\vartheta - \Theta(x,y)] ds, \label{sysenr}
\end{eqnarray}
where
\begin{eqnarray}
{\cal H}_b=&& \frac{K_s}{2} | \nabla \cdot \hat{c}(x, y)|^2 +
\frac{K_b}{2} | \nabla \times \hat{c}(x, y)|^2,\label{enrdens}\\
\sigma(\phi)=&& \sigma_0+\sum_{n=1}a_n \cos n\phi.  \label{expansion}
\end{eqnarray}
\vadjust{\vskip 0.1in
\begin{figure}
\narrowtext 
\centerline{\epsfig{file=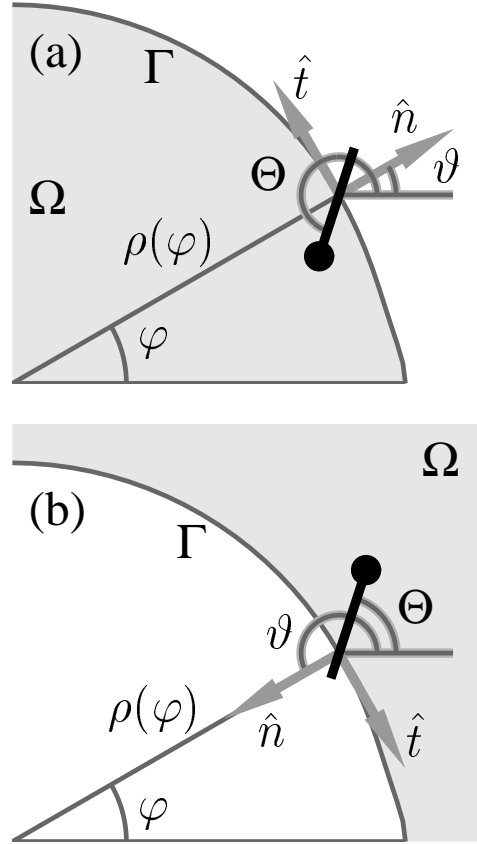, width=3in}}
\caption{
The geometry of the calculations for (a) domains and (b) bubbles
in plane-polar coordinates where the boundary $\Gamma$ is parametrized by
$\rho(\varphi)$.  The gray area is the bulk designated by $\Omega$.
$\hat{n}$ and $\hat{t}$ are
the outward normal and the tangent, respectively. $\Theta$ is the angle
between the $\hat{c}$ director and the $x$ axis and $\vartheta$ is the angle
between the outward normal of the boundary and the $x$ axis.}
\label{plst}
\end{figure}
\vskip 0.05 in}
Here, $K_s$ and $K_b$ are, respectively, the splay and bend elastic moduli, 
 and $\vartheta$ is the angle between the
outward normal of the boundary and the $x$ axis.  The quantity $\sigma_0
> 0$ is the isotropic line tension.  The first integral is over the
area, $\Omega$, of the system, while the second is over the boundary,
$\Gamma$, as indicated.  The setup of the problem in plane-polar 
coordinates is shown in Fig.~\ref{plst}. 

Minimization of the energy leads to
equations for $\Theta(x, y)$ and the bounding curve $\Gamma$.
$\Theta(x, y)$ satisfies
\begin{eqnarray}
-\nabla^2\Theta + b\left[\left(\Theta_{xx}-\Theta_{yy}\right)\cos 2
\Theta +2\Theta_{xy} \sin 2 \Theta \right.\nonumber\\
\left.+ \left(-\Theta_x^2+\Theta_y^2\right)\sin
2\Theta+2\Theta_x\Theta_y \cos 2 \Theta\right]&&= 0\label{mbulk}
\end{eqnarray}
in $\Omega$ and
\begin{eqnarray}
\kappa\Theta_n\left[1 - b \cos 2 (\vartheta-\Theta )\right]+&& \nonumber\\
\kappa b\Theta_t\sin2(\vartheta-\Theta)-\sigma^{\prime}(\vartheta-\Theta)
&&=0\label{mbc}
\end{eqnarray}
along $\Gamma$, where $\Theta_n=\hat{n}\cdot\nabla\Theta$,
$\Theta_t=\hat{t}\cdot\nabla\Theta$, $\hat{n}$ and $\hat{t}$ are,
respectively, the outward normal and tangential vectors, and
\begin{eqnarray}
\kappa=\frac{K_s+K_b}{2},\\
b=\frac{K_s-K_b}{K_s+K_b}.
\end{eqnarray}
The primes attached to functions denote derivatives, e.g., 
$\sigma^\prime(\phi)=d\sigma(\phi)/d\phi$.
The extremum equation for the bounding curve $\Gamma$, implicitly in
terms of $\Theta_n$, $\Theta_t$, and $d\vartheta/ds$, is
\begin{eqnarray}
{\cal
H}_b-\sigma^{\prime}(\vartheta-\Theta)\Theta_n-\sigma^{\prime\prime}
(\vartheta-\Theta)\Theta_t && \nonumber \\
+\left[\sigma(\vartheta-\Theta)+\sigma^{\prime\prime}(\vartheta-\Theta)\right]
\frac{d\vartheta}{ds} + \lambda&&=0,\label{mGamma}
\end{eqnarray}
where $ds$ is the length element of $\Gamma$ traversing in the
positive direction of $\Omega$ and $\lambda$ is a Lagrange multiplier
that enforces the condition of constant enclosed area.  The set of
equations, Eqs.  (\ref{mbulk}), (\ref{mbc}), and (\ref{mGamma}), are
highly nonlinear.  It appears, in general, impossible to find general
analytical solutions to this set of equations.  However, there are
full analytical solutions for special cases.

\section{Exact solutions}
\label{sec:exactsolutions}
We start with the assumption of a circular boundary.  We restrict our
considerations to isotropic elastic moduli, i.e.,  $b=0$.
Additionally, we assume that the anisotropic line tension, as given by
the expansion in Eq.~(\ref{expansion}), contains only one term, in
that $a_n=0$ for all $n\neq p$.  We will take $a_p >0$.  This is
because the texture with $a_p <0$ can be trivially obtained by
rotating all $\hat{c}(x,\,y)$ simultaneously by $(2m+1)\pi/p$,
where $m$ is an integer $\in$ $[0, p-1]$,  due to the symmetry in the
line tension. In this special case,
Eq.~(\ref{mbulk}) reduces to Laplace's equation
\begin{equation}
\nabla^2\Theta=0, \label{Laplace}
\end{equation}
and Eq.~(\ref{mbc}) in the plane-polar coordinate system becomes
\begin{eqnarray}
\kappa\Theta_\rho-\sigma^\prime(\varphi-\Theta)&=&0,\label{boundd}\\
\kappa\Theta_\rho+\sigma^\prime(\pi+\varphi-\Theta)&=&0,\label{boundb}
\end{eqnarray}
where Eq.(\ref{boundd}) applies to the case of a domain while Eq.
(\ref{boundb}) is appropriate to the case of a bubble.  In two
dimensions, the solution to Laplace's equation can be written in
general as
\begin{eqnarray}
\Theta(k,\varphi)=\frac{1}{i}[f(\mbox{e}^{k+i\varphi})-f(\mbox{e}^{k-i
\varphi})],
\label{harmonic}
\end{eqnarray}
with $f(z)$ an analytic function of $z=\mbox{e}^{k+i\varphi}$ in the
region of interest,  $\Omega$ for our case.  In the case of a
circular domain of radius $R_0$ centered at the
origin, it is shown in Appendix~\ref{samplecalc} that
\vadjust{\vskip .1in 
\begin{figure}
\centerline{\epsfig{file=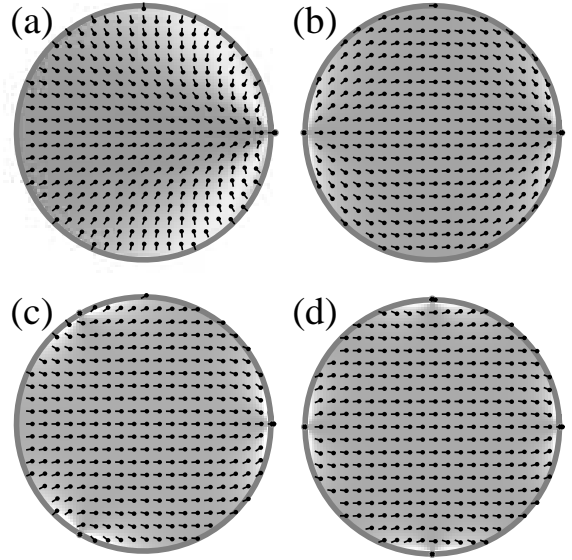, width=3in}}
\caption{The $\hat{c}$ director distribution and the BAM reflectance in a domain
computed for $\kappa=1$, $R_0=20$, and $a_p=1.6$,   where $p=1$ in (a),
$p=2$ in (b), $p=3$ in (c), and $p=4$ in (d). }
\label{boojr}
\end{figure}
\vskip .05in}
\begin{eqnarray}
\Theta_0(k, \varphi) &=&
\frac{1}{i}\left[f_0(\mbox{e}^{k+i\varphi})-f_0(\mbox{e}^{k-i\varphi})\right],\\
f_0(z) &=& \frac{1}{p}\ln(1-\alpha_p z^p), \label{boojtex}\\
\alpha_p R_0^{p} &=& -\epsilon +\sqrt{1+\epsilon^2},\label{defpos}
\end{eqnarray}
satisfy Eq.~(\ref{boundd}). We have defined here a dimensionless
parameter $\epsilon=\kappa/(p a_p R_0)$. 
Figure~\ref{boojr} illustrates such solutions for $p=1,\;2,\;3,\mbox{ and } 4$.
Also displayed in the figure on the background of each plot is a 
simulation of the image that would be obtained by Brewster angle 
microcopy (BAM).  The BAM reflectance depends on the exact experimental 
setup and the properties of the monolayer. A detailed discussion
on the computation of the BAM reflectance can be found in
Ref.~\cite{Teer}.  In the case of all simulated images presented in
Fig.~\ref{boojr} and elsewhere in this paper, the Brewster angle 
is taken to be that of water $\Theta_B=53.12^\circ$, the angle of the 
analyzer $\alpha$ is equal to $90^\circ$, the thickness of 
the monolayer is assumed to be $d=0.3\;nm$, the tilt $\Psi$ is $30^\circ$, 
the dielectric constants of the monolayer are $\epsilon_\perp=2.31$ and 
$\epsilon_\parallel=2.53$, and it is assumed that the wavelength of 
the light $\lambda=514\;nm$.
Figure~\ref{boojbd} shows the order-parameter distribution along
the boundary for the solution for $p=1$.  The plot of the order-parameter
distribution along the boundary is an effective way to examine the texture
quantitatively.

When $p=1$, the resulting texture is referred to as the
{\em boojum} texture.  It corresponds to a defect with winding
number +2 \cite{defects} lying a distance $R_B=1/\alpha_1$ from the
center of the domain.  As $\epsilon\rightarrow \infty$, the virtual
defect retreats to infinity.  As $\epsilon\rightarrow 0$,
\vadjust{\vskip .1in 
\narrowtext
\begin{figure}
\centerline{\epsfig{file=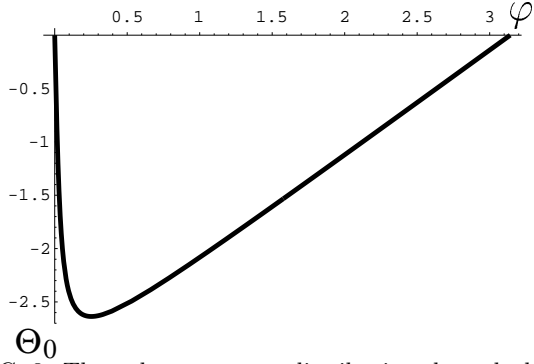, width=3in}}
\caption{The order-parameter distribution along the boundary shown as
a plot of $\Theta_0$ versus $\varphi$, where $\varphi$ is the polar
angle in the plane-polar coordinates.  The parameters are $\kappa=1$,
$R_0=20$, and $a_1=1.6$}
\label{boojbd}
\end{figure}
\vskip .05in
}
corresponding to a very strong anisotropic surface energy, or a very
large domain, the virtual defect approaches the edge of the domain.
However, the distance of the virtual singularity from the boundary of
a very large domain remains nonzero, approaching the limit $\kappa/a_1$
as $\epsilon\rightarrow\ 0$.

For the case of a bubble, instead of Eq.~(\ref{harmonic}), we make
use of
\begin{eqnarray}
\Theta(k,\varphi) =
\frac{1}{i}\left[f\left(\mbox{e}^{-k+i\varphi}\right)-f\left(\mbox{e}^{-k-i
\varphi}\right)
\right],
\end{eqnarray}
as a solution to Laplace's equation.  We find that
\begin{eqnarray}
\Theta_i(k, \varphi) &=&
\frac{1}{i}\left[f_0\left(\mbox{e}^{-k+i\varphi}\right)-f_0\left(\mbox{e}^{-k-i
\varphi}\right)\right]+\pi,\label{ibooj}\\
\frac{\alpha_p}{ R_0^{p}} &=& -\epsilon
+\sqrt{1+\epsilon^2},\label{idefpos}
\end{eqnarray}
satisfy Eq.~(\ref{boundb}) in the case of a circular bubble of
radius $R_0$ centered at the origin.  $\Theta_i$ is show in Fig.~\ref{iboojr}
for $p=1,\;2,\;3,\;\mbox{ and } 4$.
Also shown in the background of each plot
is the intensity distribution that would be recorded in a BAM image.

When $p=1$, $\Theta_i$ in Eq.~(\ref{ibooj}) can be
characterized as an inverse boojum
\vadjust{\vskip 0.1in
\narrowtext
\begin{figure}
\centerline{\epsfig{file=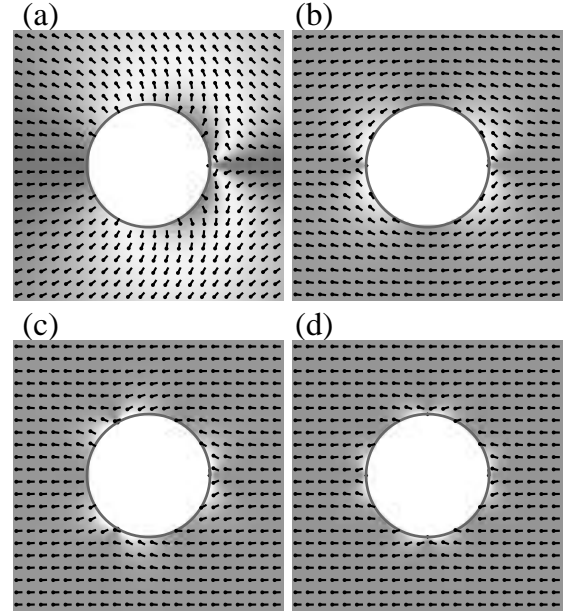, width=3in}}
\caption{The $\hat{c}$ director distribution and the BAM reflectance
for bubble computed for $\kappa=1$, $R_0=20$, $a_p=1.6$,   where $p=1$ in
(a), $p=2$ in (b), $p=3$ in (c), and $p=4$ in (d). }
\label{iboojr}
\end{figure}
\vskip 0.05in
}
texture, in that it is obtained by replacing $k$ by $-k$ in $\Theta_0$.
This corresponds to a defect with winding number $-2$ located at a
distance $R_B= \alpha_1$ from the center of the bubble.  When
$\epsilon=\infty$, the defect is at the origin.  As
$\epsilon\rightarrow 0$, it moves towards the edge of the bubble and
approaches a distance, $\kappa/a_1$, from the boundary.

Note that in the above discussion the domain and the bubble have been
\emph{assumed} to be circular.  There is no \emph{a priori} assurance
that this shape minimizes the energy of the system.

As the next step, we determine the equilibrium shape of the domain or
bubble.  Rewriting $\Gamma$ as $\rho(\varphi)\equiv \mbox{e}^{k(\varphi)}$,
that is,
in polar coordinates, we transform Eq.~(\ref{mGamma}) into
\end{multicols}
\widetext
\top{-2.8cm}
\begin{eqnarray}
\pm{\cal
H}_b\mbox{e}^k+\left\{-\sigma^{\prime}(\vartheta-\Theta)\Theta_k-
\sigma^{\prime\prime}(\vartheta-\Theta)\Theta_\varphi +
\left[\sigma^{\prime}(\vartheta-\Theta)\Theta_\varphi-\sigma^{\prime\prime}
(\vartheta-\Theta)\Theta_k\right]k^{\prime}\right.&&\nonumber\\
\left.+\left[\sigma(\vartheta-\Theta)+\sigma^{\prime\prime}(\vartheta-\Theta)
\right]\left(1-\frac{k^{\prime\prime}}{1+k^{\prime 2}}\right)
\right\}\frac{1}{\sqrt{1+k^{\prime 2}}}+\lambda&&=0,\label{polar}
\end{eqnarray}
\bottom{-2.7cm}
\begin{multicols}{2}
\hspace{-.15in}where
\begin{eqnarray}
\vartheta = \left\{\begin{array}{ll} 
\varphi-\tan^{-1}k^\prime & \mbox{for domains} \\
\pi + \varphi-\tan^{-1}k^\prime & \mbox{for bubbles.}
\end{array}\right.
\end{eqnarray}
In Eq.~(\ref{polar}), $+$ applies in the case of a domain while $-$
is appropriate to the case of a bubble.  Equation  (\ref{polar}) is a
nonlinear second-order differential equation, and there is no
indication that an analytic solution is possible.  However, for the
specific case of domain in which $b=0$ and $a_{n \neq 1}=0$ except
$a_1$, it can be verified that a circular boundary centered at the
origin is indeed a solution.
Furthermore, such a texture-boundary combination has been
shown~\cite{PettyLuben} to be a locally stable configuration.
Interestingly, a
circular boundary with the inverse boojum texture fails to satisfy
Eq.~(\ref{polar}) in the case of a bubble.
\vadjust{\vskip 0.1in
\narrowtext
\begin{figure}
\centerline{\epsfig{file=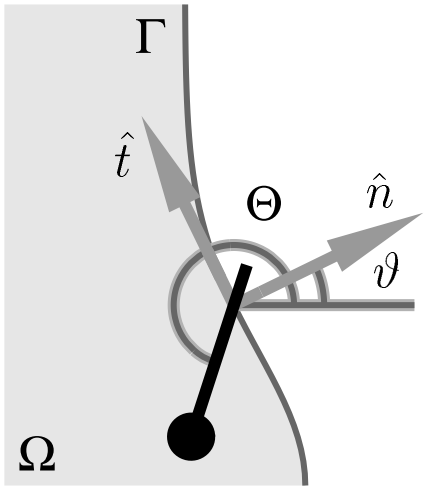, width=2.5in}}
\caption{The geometry of the calculations for domains and bubbles in 
Cartesian coordinates. The gray area is the bulk designated by
$\Omega$.
$\hat{n}$ and $\hat{t}$
are the outward normal and the tangent, respectively. $\Theta$ is the angle
made between
the $\hat{c}$ director and the $x$ axis and $\vartheta$ is the angle made
by the outward normal of the boundary and the $x$ axis. }
\label{ctst}
\end{figure}
\vskip 0.05in}

\section{Texture and Boundary Shape}
\label{sec:textureboundary}

In this section, we assume that the virtual boojum singularity lies close
to the boundary between a domain or bubble and the neighboring medium,
and we focus on the boundary in the immediate neighborhood of the
singularity.  This allows us to treat the two regions as
semi-infinite.  The anisotropic phase is taken to occupy (approximately) 
the half-space for which $x$ is negative.  The setup of 
the computation is depicted in Fig.~\ref{ctst}.  We first fix the boundary
to lie along the $y$ axis.  We then determine the texture in the
anisotropic phase when $b=0$ and all the $a_n$'s except $a_1$ are
equal to zero.  The order-parameter field $\Theta(x, y)$ will be of
the form displayed in Eq.~(\ref{harmonic}), satisfying the boundary
condition
\vadjust{\vskip 0.1in
\narrowtext
\begin{figure}
\centerline{\epsfig{file=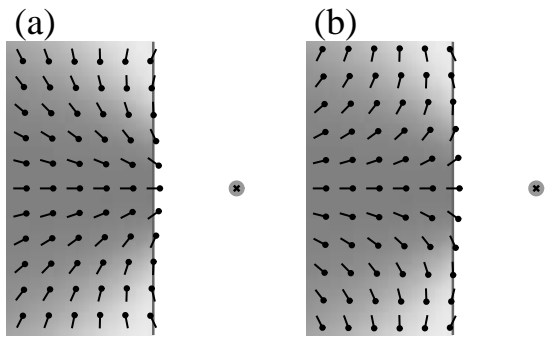,width=3in}}
\caption{The order-parameter distribution and the BAM reflectance
when the boundary $\Gamma$ is a
straight line for $\kappa=1$ and $a_1=1.6$, where (a) shows $\Theta_{C0}$
for the case of a domain while (b) shows $\Theta_{Ci}$ for the case of a 
bubble.}
\label{stbooj}
\end{figure}
\vskip 0.05in}
\begin{eqnarray}
\left[\kappa\Theta_x+a_1\sin(-\Theta)\right]_{x=0}=0.\label{cartbound}
\end{eqnarray}
This boundary condition is satisfied by the following expression:
\begin{eqnarray}
\Theta_{C0}(x, y)&=&\frac{1}{i}[f_0(x+iy)-f_0(x-iy)]\label{carttexture},
\end{eqnarray}
where $f_{0}(z)$ is as given in Eq.~(\ref{boojtex}) and
$\alpha=a_1/\kappa$.  This texture in fact corresponds to that of a
domain in polar coordinates.  Inspection of the boundary
condition Eq.~(\ref{cartbound}) leads us to another solution
$\Theta_{Ci}=-\Theta_{C0}$, which corresponds to the texture of a
bubble in cylindrical geometry.  Figures~\ref{stbooj}(a) and \ref{stbooj}(b)
show the distributions of the order parameter and the computed BAM
image in this geometry for the domain and the bubble, respectively.
The correspondence between $\Theta_{C0}$ and $\Theta_0$ for the case
of domains can be observed in Fig.~\ref{stbooj}(a) in that the
$\hat{c}$ directors tend to point towards each other around the
$x$ axis.  Figure~\ref{stbooj}(b) corresponds to Fig.~\ref{iboojr}(a)
for the case of bubbles in that the $\hat{c}$ directors fan out in the
direction of the outward normal of the boundary near the $x$ axis.

To investigate the equilibrium condition for the boundary $\Gamma$, we
parametrize $\Gamma$ in Cartesian coordinates by $x=\eta(y)$.
The equilibrium condition for $\Gamma$ can then be written as
\end{multicols}
\widetext
\top{-2.8cm}
\begin{eqnarray}
\frac{\kappa}{2}\left|\nabla\Theta\right|^2+\sigma^{\prime}(\vartheta-\Theta)
\frac{\eta^\prime\Theta_y-\Theta_x}{\sqrt{1+\eta^{\prime
2}}} +\frac{d\vartheta}{d\eta^\prime}\sigma^{\prime\prime}
(\vartheta-\Theta)\left(\Theta_y+\eta^\prime\Theta_x\right)\sqrt{1+\eta^{\prime
2}}&&\nonumber \\
-\left[\sigma(\vartheta-\Theta)+\sigma^{\prime\prime}(\vartheta-\Theta)\right]
\frac{\eta^{\prime\prime}}{\sqrt{1+\eta^{\prime
2}}^3}+\lambda&&=0,\label{equib}
\end{eqnarray}
\bottom{-2.8cm}
\begin{multicols}{2}
\hspace{-.15in}and for the boundary condition at $y=0$
\begin{eqnarray}
\eta^\prime=\frac{d\vartheta}{d\eta^\prime}\left(1+\eta^{\prime
2}\right)\frac{\sigma^\prime(\vartheta)}{\sigma(\vartheta)},\label{bcenter}
\end{eqnarray}
where $\vartheta=-\tan^{-1}\eta^\prime$ is the angle between the
outward normal $\hat{n}$ of $\Gamma$ and the $x$ axis.

A cusplike singularity occurs when $\eta^\prime(0)\neq0$.  As a
result of the symmetry of the problem, $\eta(-y)=\eta(y)$ and
$\eta^\prime(-y)=-\eta^\prime(y)$.  The possible values of
$\eta^\prime(0)$ can be obtained by solving Eq.~(\ref{bcenter}).
That $\eta^\prime(0)=0$ is a solution follows from the fact that
$\sigma^\prime(0)=0$.  In order that a nonzero $\eta^\prime(0)$
solves Eq.~(\ref{bcenter}), the slope of the right-hand side
of Eq.~(\ref{bcenter}) at the origin must be greater than unity, or
\begin{eqnarray}
\frac{d}{d\eta^\prime}\left[\frac{d\vartheta}{d\eta^\prime}
(1+\eta^{\prime2})\frac{\sigma^\prime(\vartheta)}{\sigma(\vartheta)}
\right]_{\eta^\prime=0}\ge1,
\end{eqnarray}
which leads us to the cusp condition
$\sigma(0)\;[\sigma(0)+\sigma^{\prime\prime}(0)]\le0$.  We will
exclude such a condition from our discussion, as it requires either
$|a_1|\ge\sigma_0$ or $3a_2\ge\sigma_0$.  Such conditions are
incompatible with the parameter regime on which we focus.

It can be shown that $x=0$ or the
$y$ axis is a solution to Eq.~(\ref{equib}) when $\Theta_{C0}$
is used as the texture (for the case of domain) while this is not
so for the case of the bubble, or when the texture is $\Theta_{Ci}$.
We continue to investigate
perturbatively the response to the texture when the boundary deviates
slightly from $x=0$.  Let the texture, $\Theta(x, y)$, be of the form
of Eq.~(\ref{harmonic}) with
\begin{eqnarray}
f(z)\approx f_0(z)+\xi(z)f^\prime_0(z).\label{texdist}
\end{eqnarray}
Both $\eta(y)$, the deviation of the boundary from the straight line
$x=0$, and $\xi(z)$ in Eq.~(\ref{texdist}) are small quantities.
They possess the following properties:
\begin{eqnarray}
\eta(x)=\eta(-x)\\
\xi(z)=\xi^*(z^*)
\end{eqnarray}
The boundary condition can then be expressed as
\begin{eqnarray}
\left.\kappa\left(\frac{\partial}{\partial
x}+\eta\frac{\partial^2}{\partial
x^2}-\vartheta\frac{\partial}{\partial
y}\right)\Theta\right|_{x=0}+&& \nonumber\\
\left.\rule{0in}{.2in}a_1\sin(\vartheta-\Theta)\right|_{x=\eta}&
\approx&0,
\end{eqnarray}
where $\vartheta\approx-\eta^\prime$.  To first order in $\eta,
\xi$ and their derivatives, we obtain the following equations:
\begin{eqnarray}
\frac{2\alpha y
\xi^\prime(iy)}{1+\alpha^2y^2}=-\frac{2\alpha^2y\eta(y)}{1+\alpha^2y^2}\pm
\vartheta,\label{xiexpr}
\end{eqnarray}
where $+$ corresponds to the case of domain and $-$ corresponds to the case of
bubble. The relation between $\eta$ and $\xi$ is readily derived:
\begin{eqnarray}
\xi(iy)=\int^{iy}_0\frac{1+\alpha^2y^{\prime 2}}{2\alpha
y^\prime}\left[-\frac{2\alpha^2y^\prime\eta(y^\prime)}{1+\alpha^2y^{\prime
2}}\pm\vartheta\right] dy^\prime. \label{deform}
\end{eqnarray}
Here, we distinguish the primes attached to functions which denote 
derivatives and those attached to variables within the integrals 
which indicate that they are variables of integration.  Primes will 
be used in this fashion from now on.  The function $\xi$ obtained 
from Eq.~(\ref{deform}) is defined along the $y$ axis.  We now 
analytically continue $\xi$ to the entire $x-y$ plane.  Up to this 
point, we have expressed the distortion of the texture in terms of 
a fixed boundary deviation from the exact solution given in the beginning 
of the section.  To examine the solution Eq.~(\ref{deform}), it is 
necessary to determine the textural deformation associated with 
$\eta(y)=\eta_0$ for domain.  We find
\begin{eqnarray}
\xi(z)=-\alpha\eta_0 z.
\end{eqnarray}
The resulting texture
\begin{eqnarray}
\Theta(x, y)&\approx &\frac{1}{i}[f_0(x-\eta_0+iy)-f_0(x-\eta_0-iy)]
\end{eqnarray}
is identical to the domain texture given in Eq.~(\ref{carttexture}) 
except that the position of the virtual defect is translated by an amount 
$\eta_0$ in the positive $x$ direction.

Under a small boundary distortion, $x=\eta(y)$ away from the $y$ axis,
the function $f(z)$ given in Eq.~(\ref{texdist}) and its derivative
are approximated as
\begin{eqnarray}
f(\eta+iy)&=&\ln(1-i\alpha y)-\frac{ \alpha\xi(iy)}{1-i\alpha y},\\
f^\prime(\eta+iy)&=&-\frac{\alpha}{1-i\alpha
y}-\frac{\alpha\xi^\prime(iy)}{1-i\alpha
y}-\nonumber\\
&&\qquad\frac{\alpha^2\left[\eta+\xi(iy)\right]}{(1+i\alpha y)^2},
\end{eqnarray}
where $\xi(iy)$ is given in Eq.~(\ref{xiexpr}). When these expressions
are substituted into Eq.~(\ref{equib}) we obtain, for 
$\Theta = \Theta_{C0}$, which corresponds to the case of domain,
\begin{eqnarray}
\eta^{\prime\prime}=\frac{2\delta\alpha}{1+\alpha^2
y^2}\left[\alpha\eta+\xi^\prime(iy)\right],
\label{stdom}
\end{eqnarray}
while, for $\Theta=\Theta_{Ci}$, which corresponds to the
case of bubbles,
\end{multicols}
\top{-2.8cm}
\begin{eqnarray}
\eta^{\prime\prime}=&&\delta\left[\frac{4\alpha\left(1-\alpha^2y^2\right)}
{\left(1+\alpha^2y^2\right)^2}+
\frac{2\alpha^2\left(3-10\alpha^2y^2+3\alpha^2y^4\right)}{\left(1+\alpha^2y^2
\right)^3}\eta+\frac{8i\alpha^3y\left(3-\alpha^2y^2\right)}
{\left(1+\alpha^2y^2
\right)^3}\xi(iy)+
\frac{2\alpha\left(3-\alpha^2y^2\right)}{\left(1+\alpha^2y^2\right)^2}\xi^
{\prime}(iy)\right],\label{stbub}
\end{eqnarray}
\bottom{-2.7cm}
\begin{multicols}{2}
\hspace{-.15in}where $\delta\equiv a_1/\sigma_0$.  We can see that the
right-hand side of the equation for the distorting effect of the
texture appropriate to a domain, i.e.,  Eq.~(\ref{stdom}), starts at
first order in $\eta$ while the corresponding equation, Eq.
(\ref{stbub}), for a bubble starts at zeroth order in $\eta$.  This
provides further confirmation that there is no simple inversion
symmetry between the domain and bubble.

\section{ Domains }

\label{sec:moredomain}

We have established that the boojum texture together with a circular
boundary is an exact solution for the case in which $b=0$ and
only $a_1\neq0$.  This leads us to the conclusion that nonzero $b$
and/or higher harmonics in the expansion Eq.~(\ref{expansion}) must
be present if the boundary is to be noncircular.  We now attempt to
analyze the situation in which $a_1$, $a_2$, and $b$ are all nonzero.
We do this by perturbing about the boojum solution in terms of
small parameters $b$ and $\gamma$, where $\gamma\equiv a_2/a_1$.

We note here that the sign of $a_1$ does not affect any of the
following analysis.  As has been mentioned in Sec.~\ref{sec:approach}, 
if $\Theta_0$ is an equilibrium texture for $a_1 >
0$, then $\Theta=\Theta_0+\pi$, representing a reflection of the
$\hat{c}$ directors about the $x$ axis, will be the corresponding
texture for $a_1<0$.  The equilibrium boundary is circular in both
cases.  Furthermore, contributions of $b$ and $a_2$ appear in the form
of $\Theta_z \mbox{e}^{i2\Theta}$, which is invariant under reflection of 
the $\hat{c}$ directors about the $x$ axis.  $\Theta_z$ represents
derivatives of $\Theta$ with respect to the variable $z$ which can be
$x$, $y$, or any linear combination of the two.  Hence, the effect of 
$a_2$ and $b$ is independent of the sign of $a_1$.  In the 
upcoming discussions, we will assume $a_1>0$ for convenience.  The 
inequality $\gamma > 0$ refers to the case in which the $\hat{c}$ directors 
along the boundary prefer to lie tangent to it while $\gamma <0$ applies 
when the $\hat{c}$ directors prefer to point along the normal to the
boundary, $\pm\hat{n}$.  When $b>0$, bend textures are preferred;  
splay is preferred when $b <0$.

We first find the textural response to $\gamma$ and $b$
by making use of Eq.~(\ref{harmonic}) with
\begin{eqnarray}
f(z)= f_0(z)+f_1(z).
\end{eqnarray}
It can be shown that Eq.~(\ref{mbulk}) is satisfied even with $f_1(z)=0$.
We continue to investigate the boundary condition of the texture
assuming that the bounding curve is a circle of radius $R_0$.
Equation (\ref{mbc}) requires a nonzero $f_1(z)$ satisfying
\end{multicols}
\top{-2.8cm}
\begin{eqnarray}
\kappa[z f_1^{\prime}(z)-z^{-1}f_1^{\prime}(z^*)] + 
\frac{a_1}{2}\left[\left(-\frac{z-\alpha}{1-\alpha z}-\frac{1-\alpha
z}{z-\alpha}\right)f_1(z)+\left(\frac{z-\alpha}{1-\alpha
z}+\frac{1-\alpha z}{z-\alpha}\right)f_1(z^*)\right] &&\nonumber \\
+\frac{\kappa b\alpha}{R_0}\left[\frac{1-\alpha z}{z(1-\alpha
z^*)^2}-\frac{z(1-\alpha z^*)}{(1-\alpha z)^2}\right]+
a_2\left[\left(\frac{z-\alpha}{1-\alpha
z}\right)^2-\left(\frac{1-\alpha z}{z-\alpha}\right)^2\right] &&=0.
\label{texcor}
\end{eqnarray}
In contrast to the notation used in Eqs. (\ref{boojtex}) and
(\ref{defpos}), we have redefined $z\equiv \mbox{e}^{i\varphi}$ and
absorbed $R_0$ into $\alpha\equiv R_0 \alpha_1$. Equation
(\ref{texcor}) can be separated into two equations:
\begin{eqnarray}
z f^{\prime}_1(z)-\frac{1}{2\epsilon}\left(\frac{z-\alpha}{1-\alpha
z}+\frac{1-\alpha z}{z-\alpha}\right)f_1(z)=
b\alpha\frac{z-\alpha}{z(1-\alpha
z)^2}-\frac{\gamma}{\epsilon}\left(\frac{z-\alpha}{1-\alpha
z}\right)^2
\end{eqnarray}
and an identical equation with $z$ replaced by its complex conjugate
$z^*$.  Each of these equations is solvable by standard methods.
One finds
\begin{eqnarray}
f_1(z)&=& \frac{\gamma}{\alpha (\alpha
+\epsilon)}\frac{z-\alpha}{1-\alpha z} -\frac{\epsilon(b
\alpha-2\gamma)}{\alpha + \epsilon}\frac{(z-\alpha)}{1-\alpha z}\;\/
_2F_1(1, \alpha/\epsilon + 1;\alpha/\epsilon + 2;-\alpha z),\label{texturecorr}
\end{eqnarray}
where $_2F_1(\nu, \mu;\mu+1; z)$ is a hypergeometric
function\cite{GradRyzh}.  With $f_1(z)$ included, $\Theta(x, y)$ now
satisfies Eq.~(\ref{mbulk}) up to
first order in $\gamma$ and $b$.

The full analytical solution of Eq.~(\ref{polar}) is difficult when
$\gamma$ and $b$ are both nonzero.  However, one can attempt a
solution as an expansion in the small parameters $\gamma$, $b$, $k$,
and $k^{\prime}$.  We recall the definition of $\rho(\varphi)
\equiv \mbox{e}^{k(\varphi)}\approx 1+k(\varphi)$.  If one ignores terms beyond
first order in these quantities, it is possible to solve for the bounding
curve, $\Gamma$, analytically.  The algebraic manipulations are dramatically
simplified if we further approximate $\Theta_n\approx\Theta_k/R_0$
and $\Theta_t\approx\Theta_\varphi/R_0$, which is equivalent to 
neglecting the first-order contributions of $k^\prime(\varphi)$.  
The error of the analysis is
then of order $\sim O(\delta k^\prime)$, where $\delta\equiv a_1/\sigma_0$,
which has been defined earlier in Sec.~\ref{sec:textureboundary}.
The equation for the boundary $\Gamma$,
\begin{eqnarray}
{\cal
H}_bR_0+\left\{-\sigma^{\prime}(\varphi-\Theta)\Theta_k-\sigma^{\prime\prime}
(\varphi-\Theta)\Theta_\varphi
\right.  && \nonumber \\
\left.
+\left[\sigma(\varphi-\Theta)+\sigma^{\prime\prime}(\varphi-\Theta)\right]
\left(1-k^{\prime\prime}\right)\right\}
+ \lambda&&=0,\label{domsg}
\end{eqnarray}
can then be reduced to
\begin{eqnarray}
k^{\prime\prime}(\varphi)=k_1^{\prime\prime}(\varphi)+k_2^{\prime\prime}
(\varphi)+k_3^{\prime\prime}(\varphi),\label{kpp}
\end{eqnarray} where $k^{\prime\prime}$ has been separated into various
components, as
shown below, for convenience,
\begin{eqnarray}
k_1^{\prime\prime}(\varphi)&=&\frac{\delta b\alpha^2(-z +\alpha
+\epsilon)}{(1-\alpha z)^2}+ \mbox{c.c.},\\
k_2^{\prime\prime}(\varphi)&=&-\delta\alpha\gamma\left(\frac{z}{1-\alpha
z}+\frac{3}{z-\alpha}+\frac{3}{2\alpha}-\frac{1}{\epsilon
z}\right)\left(\frac{z-\alpha}{1-\alpha z}\right)^2 + \mbox{c.c.},\\
k_3^{\prime\prime}(\varphi)&=& \delta\alpha\left[\frac{1}{1-\alpha
z}-\frac{z}{z-\alpha}-\frac{1}{2\epsilon
z}\left(\frac{z-\alpha}{1-\alpha z}+\frac{1-\alpha
z}{z-\alpha}\right)\right]f_1(z) + \mbox{c.c.}
\end{eqnarray}
$k_1(\varphi)$ consists of terms that depend on $b$,  
$k_2(\varphi)$ contains terms that depend on $\gamma$, and 
$k_3(\varphi)$ has terms that depend on both $b$ and 
$\gamma$ through the textural correction $f_1$ given in Eq.~
(\ref{texturecorr}).
The functions $k^\prime(\varphi)$ and $k(\varphi)$ can
then be obtained by integrating Eq.~(\ref{kpp})
with respect to $\varphi$.  They are
\begin{eqnarray}
k_1^\prime(\varphi)&=&\frac{\delta
b\alpha^2}{i}\left[-(\epsilon+\alpha)\ln(1-\alpha
z)+\frac{1-\alpha^2-\alpha\epsilon}{\alpha(\alpha z-1)}\right] -
\mbox{c.c.},\\
k_2^\prime(\varphi)&=&-\frac{\delta\alpha\gamma}{i}\left\{\left[\frac{1}{2\alpha
}+\frac{3\alpha^2}{2}-\frac{1-\alpha^4}{\epsilon}\right]\ln(1-\alpha
z) +\right.  \nonumber \\
&& \left.
\left[\frac{1}{2\alpha}-\frac{3\alpha}{2}+\frac{1-\alpha^2}{\epsilon}\right]
\frac{1-\alpha^2}{\alpha
z-1} + \frac{\left(1-\alpha^2\right)^2}{2\alpha(\alpha z-1)^2}\right\}
- \mbox{c.c.},\\
k_3^\prime(\varphi)&=&-
\frac{\delta\alpha}{i\epsilon}\left(b\alpha-\frac{2\gamma}{\delta}\right)
g(\alpha z)- \mbox{c.c.},
\end{eqnarray}
where
\begin{eqnarray}
g(z)&=&-\frac{\epsilon}{\alpha+\epsilon}\int^{z}\frac{_2F_1\left(1,
\alpha/\epsilon+1;\alpha/\epsilon+2; -z^\prime\right)}{z^\prime}
dz^\prime.\label{gdef}
\end{eqnarray}
One more integration yields
\begin{eqnarray}
k_1(\varphi)&=&\delta
b\alpha^2\left[\left(\epsilon-\frac{1}{\alpha}+\alpha\right)\ln(1-\alpha
z)-(\epsilon+\alpha)\mbox{Li}_2(\alpha z) \right] + \mbox{c.c.},\\
k_2(\varphi)&=&\frac{\delta\gamma}{\alpha}\left\{-\left[\frac{2}{\alpha}+
\frac{3\alpha^3}{2}-\frac{1-\alpha^4}{\epsilon}\right]\mbox{Li}_2(\alpha
z)\right. + \nonumber \\
&&\left.  \left[-\alpha(1-\alpha^2)+ \frac{\left(1-\alpha^2 \right)^2
}{\epsilon}\right]\ln(1-\alpha
z)-\frac{\left(1-\alpha^2\right)^2}{2\alpha(\alpha z -1)}\right\} +
\mbox{c.c.},\\
k_3(\varphi)&=& -
\frac{\delta\alpha}{\epsilon}\left(b\alpha-\frac{2\gamma}{\delta}\right)
h(\alpha z)+ \mbox{c.c.},\end{eqnarray}
where
\begin{eqnarray}
h(z)=\int^z g(z^\prime)/z^\prime dz^\prime\label{hdef}
\end{eqnarray}
and Li$_2(z)$ is the polylogarithmic
function defined as
\begin{eqnarray}
\mbox{Li}_n(x)&&=\sum_{k=1}^\infty\frac{x^k}{k^n}.
\end{eqnarray}
Numerical integrations can be utilized for the evaluation of $k(\varphi)$.
However, as we can see from the following,
\begin{eqnarray}
\int^z_0\frac{z^{\prime\alpha/\epsilon}}{1-\alpha
z^\prime}dz^\prime=-\frac{\epsilon
z^{\alpha/\epsilon+1}}{\alpha+\epsilon}\/_2F_1(1,
\alpha/\epsilon+1;\alpha/\epsilon+2;-\alpha z), \label{hypfunc}
\end{eqnarray}
 the integrand oscillates strongly as $\alpha/\epsilon\rightarrow\infty$ 
as a result of the factor $z^{\prime\alpha/\epsilon}$ and numerical
integrations become inefficient.  Further observation reveals that $g(z)$
and $h(z)$ can be evaluated analytically if $\alpha/\epsilon \equiv n$ is an
integer.  Equation  (\ref{hypfunc}) simplifies as follows:
\begin{eqnarray}
\int^z_0\frac{z^{\prime n}}{1-\alpha
z^\prime}dz^\prime=-\alpha^{-n-1}\left[\sum^n_{i=1}\frac{(\alpha
z)^i}{i}+\ln(1-\alpha z)\right].
\end{eqnarray}
When the above simplification is substituted into Eq.~(\ref{gdef}),
the integration can be performed and yields the analytic form of $g(z)$:
\begin{eqnarray}
g(z)=&&\sum_{k=1}^n\left[-\frac{1}{n-k+1}+\frac{1}{n+1}\right]\frac{1}{k
z^k}+\frac{\ln(1-z)}{n+1}\left[1-\frac{1}{z^{n+1}}\right].
\end{eqnarray}
$h(z)$ can be evaluated analytically in the same manner and we get
\begin{eqnarray}
h(z)=&&\sum_{k=1}^n\frac{1}{(n-k+1)k^2z^k}-\sum_{k=1}^n\left[\frac{1}{(n+1)^2k}+
\frac{1}{(n+1)k^2}\right]\frac{1}{z^k}\nonumber\\
&&+\frac{\ln(1-z)}{(n+1)^2z^{n+1}}-\frac{\ln(1-z)}{(n+1)^2}-
\frac{\mbox{Li}_2(z)}{n+1}.
\end{eqnarray}
\bottom{-2.7cm}
\begin{multicols}{2}

The boundary of the domain $\displaystyle{k(\varphi)=\sum_{j=1}^3
k_j(\varphi)}$ is
smooth and has continuous derivatives.  Typical $k(\varphi)$,
$k^\prime(\varphi)$, and $k^{\prime\prime}(\varphi)$'s are shown in
Fig.~\ref{kplt}(a).  The corresponding boundary $\Gamma$ parametrized
by $\rho(\varphi)=1+k(\varphi)$ is depicted in Fig.~\ref{kplt}(b).
We have thus arrived at an approximate expression for $\Gamma$ as a
function of the line-tension anisotropy coefficient $\gamma$ and the elastic
anisotropy coefficient $b$.  This expression is useful when we are
interested in the response of $\Gamma$ for small values of the these 
anisotropic parameters.

We first examine the boundary response to $\gamma$ while keeping
$b=0$.  We find indentations and protruding features on the domain
boundary for $\gamma<0$ and $\gamma>0$, respectively.  The progressive
change of the boundary response when $a_2$ changes from $-0.5$ to
$0.5$ is illustrated in Fig.~\ref{anadp}.  The results are in
qualitative agreement with those presented in Ref.~\cite{GalaFour}.
We have also examined the dependence of the boundary at fixed $\gamma$
on domain size, $R_0$.  Figure~\ref{anrdp} show boundaries for domains
of sizes $R_0=0.2$ to $10$.  When  $R_0<1$, the domains appear slightly 
flattened and elongated if $\gamma > 0$.  This is
in accord with the intuitive notion that the second-harmonic term in
the line tension becomes important as the variation in the texture
vanishes, i.e., in the limit that the order parameter is uniform.
As the domain gets larger, the
boojum singularity moves closer to the edge of the domain and the boundary
correction moves towards the axis connecting the center of the domain
and the boojum.  Cusplike features start to appear when the domain is
larger than a ``threshold'' size, $R_0=1$ for the domains in
Fig.~\ref{anrdp}.   We note that we have used large values of $\gamma\le0.3$
to illustrate the nontrivial boundary that we have obtained for the
domains.  It has been numerically verified \cite{numerics} that
the qualitative behavior of the boundary response is indeed preserved up
to much larger values of $\gamma$.

It has been shown in Sec.~\ref{sec:textureboundary} that the
boundaries of the domains are strictly smooth and continuous in the
parameter regime upon which we focus.  We are, however, able to find
domains with cusplike features in the context of the perturbative
analysis described in this paper.  Such domains can be characterized
by an {\em excluded angle} $\Psi_0$ defined in Fig.~\ref{andenf}(a).
Domains with boundaries that resemble those with cusplike features
are observed experimentally.  The domain-size dependence of $\Psi_0$ is
shown in Fig.~\ref{andenf}(b)~\cite{FangTeer}.  There is no rigorous
mathematical definition of $\Psi_0$ for a continuous boundary.  It is,
nevertheless, possible to devise a systematic way of identifying such
an excluded angle for a smooth boundary.  One first evaluates
$\Psi\equiv-2 \tan^{-1}dx/dy$ along the boundary.  The value of $\Psi$
at the straightest part of the boundary, which is indicated by
$d^2x/dy^2\rightarrow 0$, is a likely candidate for $\Psi_0$.
Figures~\ref{dendet}(a) and \ref{dendet}(b) show the plot of $\Psi$ and
$d^2x/dy^2$ versus $\varphi$, respectively.  The values of $\Psi$ in
the plateau region in Fig.~\ref{dendet}(a) represent the range in
which the measured excluded angles are likely to fall.  These values
of $\Psi$ are found in the region near the $\Psi$ axis of the
plot of $\Psi$ versus $d^2x/dy^2$ shown in Fig.~\ref{dendet}(c).  A
plot of $\Psi$ versus $I\equiv I_0 \exp[-(d^2x/dy^2)^2]$, as shown in
Fig.~\ref{dendet}(d), highlights the range of the values of $\Psi$
that is most likely to contain the measured excluded angle.
Figure~\ref{dendet}(e) displays $I$ as the intensity (inverted, in
that the brightest corresponds to the smallest value of I).
Figure~\ref{dendet}(f) shows the value of $\Psi$ at which $I=I_{\mbox{max}}$
by the dark line and the region in which $I>I_{\mbox{max}}/2$, or the
full-width-at-half-maximum (FWHM), by the gray band.  
Figures~\ref{dendet}(e) and \ref{dendet}(f) are useful for describing the
selection process by which one is led to the most likely values of the
excluded angle.  Figures~\ref{anden}(a) and \ref{anden}(b) illustrate such 
plots. The experimental result is superposed in Fig.~\ref{anden}(b). The
parameters are adjusted in order to obtain a by-eye fit.  The
values of the parameters are $\kappa/a_1=4\;\mu$m, $\delta=0.4$, and
$\gamma=0.5$

The perturbative analysis generates results that are in good agreement with
experimental
\vadjust{\vskip 0.1in
\narrowtext
\begin{figure}
\centerline{\epsfig{file=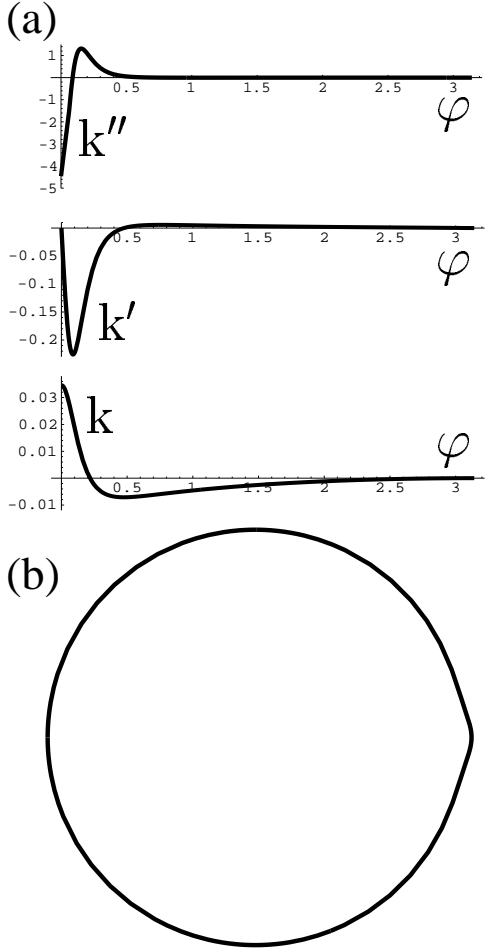,width=2.75in}}
\caption{(a) Plot of $k^{\prime\prime}(\varphi)$, $k^\prime(\varphi)$,
and $k(\varphi)$ for $\kappa=1$, $R_0=5$, $\sigma_0=4$, $a_1=1.6$, and
$a_2$=0.5.
(b) The corresponding domain shape $\Gamma$ parametrized as
$\rho(\varphi)\approx 1+k(\varphi)$.}
\label{kplt}
\end{figure}
\vskip 0.05in
}
observations~\cite{FangTeer} for large domains.  It has also captured
qualitatively the essential features, namely the onset and the
maximum of the domain size dependence of the excluded angle.  
As displayed in Fig.~\ref{anden}(b), the maximum
and the onset of $\Psi_0$ are quantitatively different in the 
perturbative analysis and in the experimental
data in the intermediate $R_0$ regime.  In particular,
the experimental maximum of $\Psi_0$ cannot be obtained from the analysis
even though $\gamma=0.5$ has been used. We shall 
defer the discussion on this to the end of this section after 
elaborating on the effect of the elastic anisotropy $b$. We also note
that $\gamma=0.5$ is very large as a perturbative parameter.  Although
there is no {\em a priori} guarantee of the accuracy of the results,
it is evident from our numerical studies\cite{numerics} that the
qualitative behavior of the boundary as a function of the domain size
is preserved in the perturbative analysis up to at least $\gamma=0.5$.

\begin{figure}
\centerline{\epsfig{file=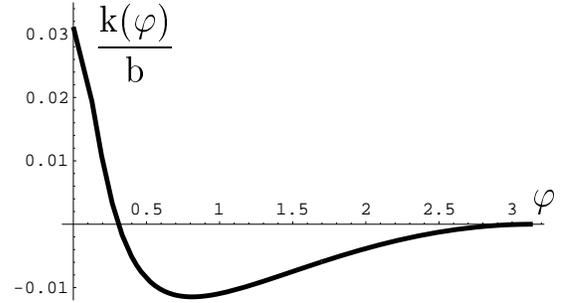,width=3in}}
\caption{Plot of $k(\varphi)/b$ for $\kappa=1$, $R_0=5$, $\sigma_0=4$,
$a_1=1.6$, and $a_2=0$. The maximum of $k(\varphi)$ at $\varphi=0$ implies a
protruding correction when $b>0$.}
\label{kbeta}
\end{figure}

We now proceed to examine the effect of $b$ on the boundary $\Gamma$.
The coefficient of the anisotropic line tension $\gamma$ is  
kept at zero and the boundary response is proportional to $b$.
Figure~\ref{kbeta} shows the plot of 
\vadjust{\vskip 0.1in
\narrowtext
\begin{figure}
\centerline{\epsfig{file=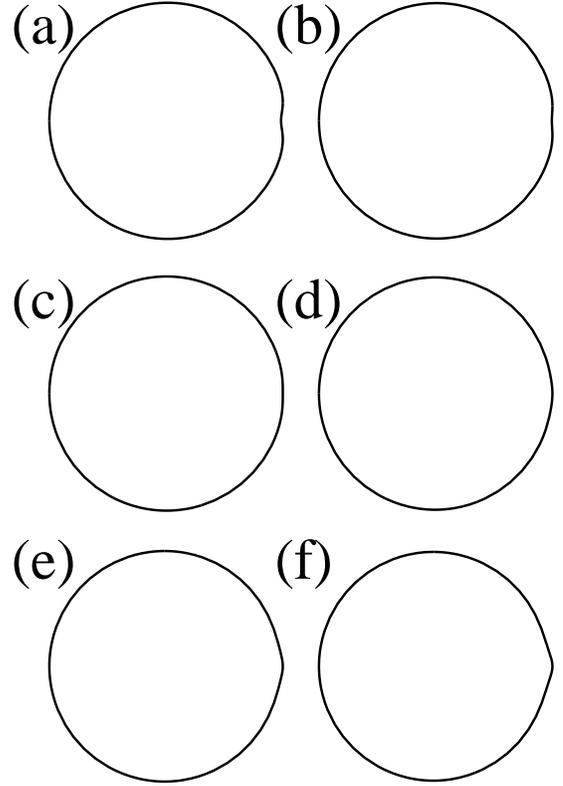,width=3in}}
\caption{Domain shapes computed for $\kappa=1$, $R_0=5$,  $\sigma_0=4$,
$a_1=1.6$. and (a) $a_2=-0.5$, (b) $a_2=-0.3$, (c) $a_2=-0.1$,
(d) $a_2=0.1$, (e) $a_2=0.3$, and (f) $a_2=0.5$.}
\label{anadp}
\end{figure}
\vskip 0.05in
}
\vadjust{\vskip 0.1in
\narrowtext
\begin{figure}
\centerline{\epsfig{file=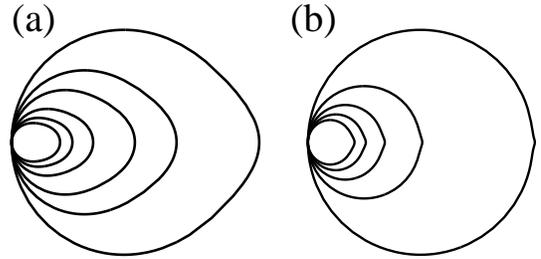,width=3in}}
\caption{Domain shapes computed for $\kappa=1$, $\sigma_0=4$, $a_1=1.6$,
$a_2=0.6$, and
(a) $R_0=0.2, 0.25, 0.33, 0.5, 1$ and (b) $R_0=2, 2.5, 3.3, 5, 10$.}
\label{anrdp}
\end{figure}
\vskip 0.05in
}
$k(\varphi)/b$.  In contrast to
the results obtained by Galatola and Fournier\cite{GalaFour}, the
boundary acquires a denting correction when $b<0$, indicated by a
maximum in $k(\varphi)/b$ at $\varphi=0$.  This perturbative result is
confirmed for small values of $b$ (=0.1) by our numerical studies
\cite{future}.  At such small values of $b$, the boundary is
practically circular.  We shall restrict our discussion to the
effect on $\Gamma$ of small values of $b$, in that higher-order
corrections of $b$, which are not taken into account in this first-order
perturbative analysis, change the qualitative behavior of the
\vadjust{\vskip 0.1in
\narrowtext
\begin{figure}
\centerline{\epsfig{file=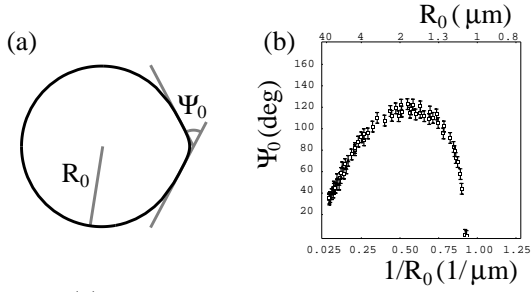,width=3in}}
\caption{(a) The definition of the excluded angle $\Psi_0$. (b)
Experimental measurements of the domain-size dependence of $\Psi_0$
observed in $L_2$ domains surrounded by LE phase taken from Ref. [5].}
\label{andenf}
\end{figure}
\vskip 0.05in}
\vadjust{\vskip 0.1in
\narrowtext
\begin{figure}
\centerline{\epsfig{file=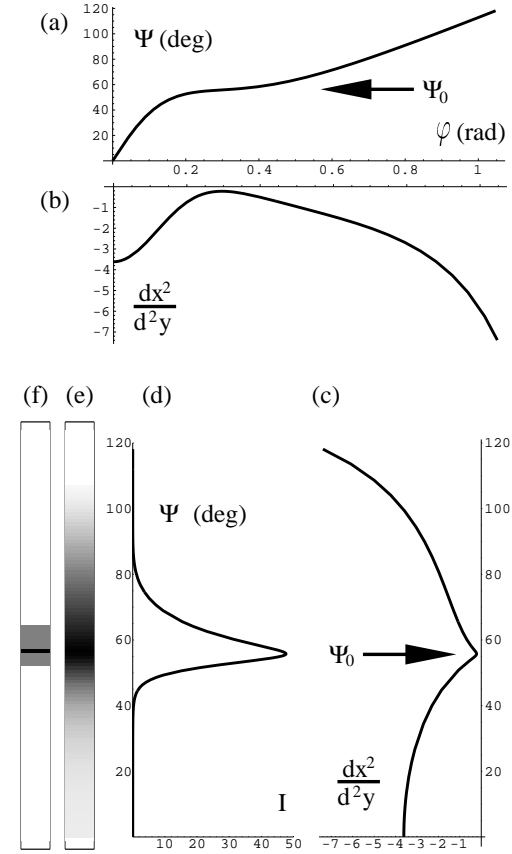,width=3in}}
\caption{(a) Plot of $\Psi\equiv-2\tan^{-1}dx/dy$ versus $\varphi$ for
$\kappa=1$, $R_0=5$, $\sigma_0=4$,
$a_1=1.6$, and $a_2=0.6$. (b) Plot of $d^2x/dy^2$ versus $\varphi$ for the same
parameter.  (c) Plot of $\Psi$ versus $d^2x/dy^2$. (d) Plot of $\Psi$
versus $I\equiv\mbox{exp}[-(d^2x/dy^2)^2]$.
(e) Density plot of $I$ as $\Psi$ for a single $R_0$.  (f) The dark line
marks the maximum $I_{\mbox{max}}$ of $I$ while the
gray region shows the range of $\Psi$ in which $I>I_{\mbox{max}}/2$. }
\label{dendet}
\end{figure}
\vskip 0.05in}
boundary response, as reflected by our numerical studies
\cite{numerics}.

%\vadjust{\vskip 0.1in
\narrowtext
\begin{figure}
\centerline{\epsfig{file=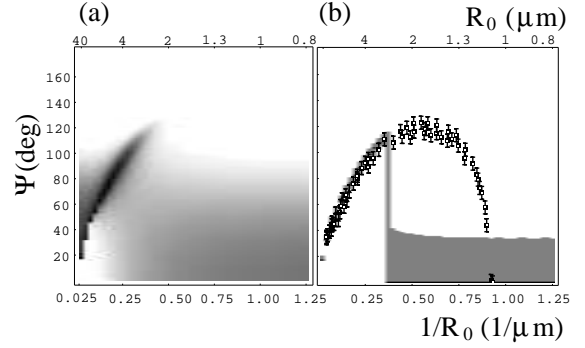,width=3in}}
\caption{(a) Density plot of $I$ as a function of $\Psi$ and $R_0$.
(b) Plot of $I_{\mbox{max}}$ and the region in which $I>I_{\mbox{max}}/2$ as a function
of $\Psi$ and $R_0$.  Superimposed are the experimental data
shown in Fig.~\ref{andenf}(b) with parameters $\kappa/a_1=4\;\mu$m, $\delta=0.4$,
and $\gamma=0.5$.}
\label{anden}
\end{figure}
%\vskip 0.05in
%}
%\vadjust{\vskip 0.1in
\narrowtext
%\vskip 0.05in
%}

As for the role of $b$ in the interpretation of the experimental
observations, we conclude in our numerical studies \cite{numerics}
that a nonzero value of $b$ cannot be solely responsible for the
protruding features, and hence the excluded-angle measurements.  Large
values of $\gamma\approx0.5$ are required to produce excluded angles
whose maximum approaches the largest value of experimentally measured
excluded angles.  At sufficiently large values of $\gamma$, we have
found that the behavior of the excluded angle is
qualitatively unchanged when the anisotropy parameter $b$ is varied
from $-0.5$ to 0.5.  Although the validity of the perturbative analysis
at $b>0.1$ is questionable, the relative magnitude of the correction
to the boundary of $b$ is much smaller than that of $\gamma$.  This is
further verified by numerical studies\cite{future}.

We have thus demonstrated, within our first-order perturbative
analysis, that the line-tension anisotropy $\gamma$ can give rise to
the indentations and protruding features of the domain boundaries that
have been experimentally observed \cite{cigar}.  Our results on the
boundary response to $\gamma$ are in qualitative agreement with prior
results\cite{GalaFour,numerics}.  Although our investigation of the
$b$ dependence of the boundary does not provide us with dependable
results for large values of $b$, it supports the assertion that the
textural correction is an important contribution to the boundary response.
Further analysis of the available experimental data on
the excluded angles leads to the conclusion that $b$ has little effect
on the boundary of the domains with protrusions.  We shall confine our
conclusions to small values of $|b|<0.1$, although large values of
$b\sim0.8$ do lead to interesting domain shapes.  Discussions of the
domain boundaries at large values of $b$ will be presented in a
forthcoming article \cite{future}, and some of the results have been
briefly presented in Ref.  \cite{numerics}.

We obtain good agreement between the results of the perturbative 
analysis and the experimental observations on the $\Psi_0$ dependence
on the domain radius $R_0$ in the large-$R_0$ regime.
In the intermediate-$R_0$ regime,  the discrepancy is not resolved, 
even in our numerical studies~\cite{future}.  The mismatch could 
possibly be attributed to the fact that our simple elastic theory 
does not describe accurately the actual complex monolayer.  
In the perturbative analysis performed in this
work,  the parameters are restricted to a region in which 
$\sigma(\varphi)+\sigma^{\prime\prime}(\varphi)$ is always 
greater than zero.  Furthermore, it is generally known that
the dipolar interactions between the surfactant molecules in the
monolayer are important.  The current model does not take into 
account such interactions. It has been discovered in a recent 
experimental study~\cite{Tabe} that the tilt is not always uniform,
especially in the region around a point defect.  The contribution
of variation in tilt may not be significant in terms of 
accounting for the discrepancy in the intermediate-sized domain 
regime.  It does become important in the large-$R_0$ regime
when the virtual singularity approaches the domain boundary
and the texture acquires a rapid variation in the neighborhood
of the virtual singularity.

\section{Bubbles}
\label{sec:bubbles}
It has been shown in Secs.~\ref{sec:exactsolutions} and
~\ref{sec:textureboundary} that there is no straightforward inversion
symmetry between the domains and the bubbles.  In contrast to the case
of the domain, it is not necessary to introduce anisotropic parameters
other than $a_1$.  The inverse boojum texture $\Theta_i$ given in Eq.
(\ref{ibooj}) and Eq.~(\ref{idefpos}) satisfies the equilibrium
condition for a circular bubble for the case in which the
$\hat{c}$ directors favor pointing into the bulk or $a_1>0$.
Substituting $\Theta_i$ into the equilibrium condition for the
boundary $\Gamma$ Eq.~(\ref{polar}), one finds that a circular boundary
does not satisfy the equation.  By perturbing about the circular
boundary in terms of a small parameter $\delta\equiv a_1/\sigma_0$, we
arrive at an equation similar to Eq.~(\ref{kpp}),
\begin{eqnarray}
k^{\prime\prime}(\varphi)&=&-2\alpha^2\delta\epsilon\frac{z^2}{(1-\alpha
z)^2}+\mbox{c.c.}
\end{eqnarray}
Again---see Eqs (\ref{ibooj}) and (\ref{idefpos})---we have redefined
$z\equiv \mbox{e}^{i\varphi}$ and $\alpha\equiv \alpha_1/R_0$.  Following the
same procedure as for the case of a domain, we find for
$k^\prime(\varphi)$ and $k(\varphi)$
\vadjust{\vskip 0.1in
\narrowtext
\begin{figure}
\centerline{\epsfig{file=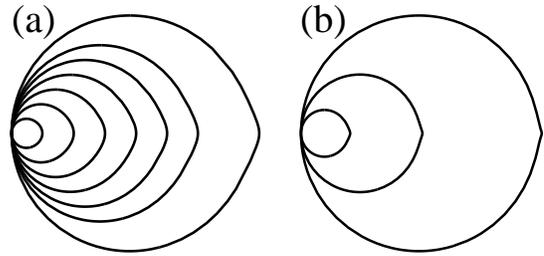,width=3in}}
\caption{Bubble shapes computed for $\kappa=0.16$, $\sigma_0=1$, $a_1=0.16$,
 and (a) $R_0=1, 2, 3, 4, 5, 6, 8$ and (b) $R_0=8, 20, 40$.}
\label{bbrdp}
\end{figure}
\vskip 0.05in}
\vadjust{\vskip 0.1in
\begin{figure}
\narrowtext
\centerline{\epsfig{file=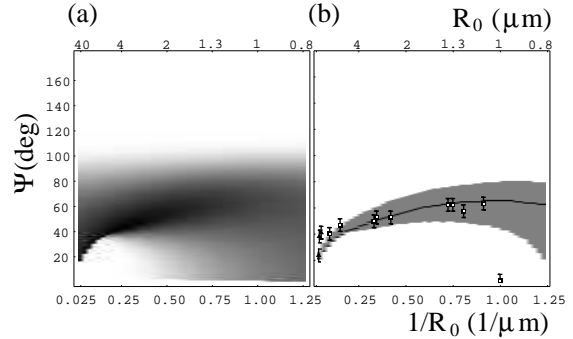,width=3in}}
\caption{
(a) Plot of $I$ as a function of $\Psi$ and $R_0$.  (b) Plot
of $I_{\mbox{max}}$ and the region in which $I>I_{\mbox{max}}/2$ as a function of
$\Psi$ and $R_0$.  Superimposed are the experimental observations 
of gaseous bubbles in $L_2$ phase.  The experimental
data have appeared in Ref. [5].  The parameters for the by-eye fit
are $\kappa/a_1=0.4\;\mu$m and $\delta=0.16.$}
\label{bbden}
\end{figure}
\vskip 0.05in}
\begin{eqnarray}
k^\prime(\varphi)&=&-\frac{2\delta\epsilon}{i}\left[\frac{1}{1-\alpha
z}+\ln(1-\alpha z)\right]-\mbox{c.c.},\\
k(\varphi) &=&-2\delta\epsilon\left[\ln(1-\alpha z)+\mbox{Li}_2(\alpha
z)\right] +\mbox{c.c.}\label{kbubble}
\end{eqnarray}
It should be kept in mind that the above discussion refers to the case
$a_1>0$.  The results for $a_1<0$ are {\em not} obtained by a
simple sign reversal of $a_1$ in Eq.~(\ref{kbubble}).  Appropriate
changes in the texture and definition of $\epsilon$, which is
given as $\kappa/(|a_1|R_0)$, must be taken into account.
The details of the calculations are presented in
Appendix~\ref{samplecalc}.  The final bounding curve for the bubble
depends only on the magnitude $|a_1|$.  We have derived expressions
for $\Gamma$ for the cases where there is only an $a_1$ contribution in
the line tension.

One can utilize the results to investigate the bubble-size dependence
of the boundary shapes.  Figure~\ref{bbrdp} shows the shapes for the
bubbles of sizes $R_0= 0.2$ to $10$.  Very small bubbles $R_0\ll1$ are
nearly circular, as are very small domains.  Cusplike features start to
appear when the bubble is larger than a ``threshold'' size, $R_0=1$
for the bubbles shown in Fig.~\ref{bbrdp}.  Similar analysis of the
excluded angle to that for the domain presented in
Sec.~\ref{sec:moredomain} can be carried out.  Figure~\ref{bbden}
shows  plots of $\Psi$ versus $R_0$ corresponding to those in
Fig.~\ref{anden}.  Experimental measurements~\cite{FangTeer} of the
cusp angle for the bubble are superposed in Fig.~\ref{bbden}(b).
A by-eye fit can be obtained with parameters $\kappa/a_1=0.4\;\mu$m
and $\delta=0.16$.  We find fairly reasonable agreement between the
theory and experimental observations.
We remark that the apparent mismatch between the theory and 
the experimental data point at $R_0=1\;\mu$m,  which has
been shown to match the explicit measurement on the 
calculated bubble boundary in Ref. \cite{FangTeer},  may be
the result of the inadequacy in qualifying the excluded
angle $\Psi_0$ for bubbles with $R_0<1\;\mu$m using the
FWHM of $I$ shown in Fig. \ref{dendet}(f).

As compared to the parameters obtained for the domains in 
Sec.~\ref{sec:moredomain}, which are $\kappa/a_1=4\;\mu$m, $\delta=0.4$, 
and $\gamma = 0.5$, the value of $\kappa/a_1$ for the case of 
bubbles is an order of magnitude smaller than that for the case 
of domains.  Noting the fact that the data for the
bubbles are obtained at the $L_2/G$ coexistence region and 
those of the domains are measured when the $L_2$ domains are 
surrounded by the $LE$ phase~\cite{FangTeer}, the comparison of 
$\kappa/a_1$ between the domains and the bubbles is indeed in 
accord with the intuitive sense that $\kappa$ of the $L_2$ domains
should not vary significantly while $a_1$ at the $L_2/G$ interface
is much larger than that at the $L_2/LE$ boundary.  The $\delta$'s
are of the same order of magnitude and there is no corresponding
$\gamma$ in the case of bubbles.  The result of the perturbative
analysis is consistent between the domains and the bubbles.

\section{Thermal Fluctuations}
\label{sec:thermal}
\narrowtext 
The analysis presented in the earlier sections is in the mean-field 
approximation; thermal fluctuations are ignored.  In this section, 
we examine the effect of thermal fluctuations and its implications
to the computation that has been carried out.
The effect of fluctuations can be assessed by utilizing a mapping
between the statistical mechanics of the order-parameter fluctuations
in this system and the behavior of a two-dimensional Coulomb
gas\cite{HP}. Consider the Hamiltonian of the
form of Eq.~(\ref{sysenr}), with $b=0$ and $a_n=0$ for all $n\neq p$.
For a system with circular boundary of radius $R_0$, one has
\begin{eqnarray}
H^\prime[\Theta]=&&H[\Theta]-2\pi\sigma_0R_0.
\end{eqnarray}
the prime here distinguishes the free energy from the one defined
in Eq.~(\ref{sysenr}).  The prime is dropped from now on for convenience.
The partition function can be written as
\begin{eqnarray}
Z(a_p)&=& Z(0) \left\langle\exp \left[-\beta a_p\oint \cos
p(\varphi-\Theta)ds\right]\right\rangle_0,\label{partitionfunction}
\end{eqnarray}
where $\langle{\cal O}\rangle_0$ denotes the thermal average with
respect to the Hamiltonian without the boundary term given below,
\begin{eqnarray}
\langle{\cal O}\rangle_0=\frac{\int{\cal D}\Theta{\cal
O}\exp(-\frac{\beta\kappa}{2}\int dA|\nabla\Theta|^2)}{\int{\cal
D}\Theta\exp(-\frac{\beta\kappa}{2}\int dA|\nabla\Theta|^2)}.
\end{eqnarray}

We denote $\Theta_b(\varphi)\equiv\Theta(R_0\cos\varphi, R_0\sin\varphi)$
as the values of $\Theta$ on the boundary of the circular domain.
The following correlation function can be evaluated\cite{Saleur},
\begin{eqnarray}
\langle\Theta_b(\varphi)\Theta_b(\varphi^\prime)\rangle_0
&=&-\Delta\ln 2\left|\sin\frac{\varphi-\varphi^\prime}{2}\right|,
\end{eqnarray} where $\Delta\equiv 1/2\pi\beta\kappa$.

To evaluate the full partition function Eq.~(\ref{partitionfunction}),
we Taylor expand the exponent as
\end{multicols}
\top{-2.8cm}
\widetext
\begin{eqnarray}
\frac{Z(a_p)}{Z(0)}=\sum_{n=0}^{\infty}\left(-\beta
a_p\right)^n\frac{1}{n!}\left\langle\prod_{i=1}^{n}\left[\oint\frac{ds_i}{a}
\cos p(\varphi_i-\Theta_i)\right]\right\rangle_0,\label{pf2}
\end{eqnarray}
where we have added  an index $i$ to distinguish the various cosine terms,
denoted $\Theta_i\equiv\Theta(R_0 \cos\varphi_i,R_0 \sin\varphi_i)$ for
convenience,
and introduced a microscopic length scale $a$ in the denominator in $ds$ to
make the
total integration dimensionless. We use
$\cos\phi=\left(\mbox{e}^{i\phi}+\mbox{e}^
{-i\phi}\right)/2$ and then expand
the products of the cosine terms,
\begin{eqnarray}
\frac{Z(a_p)}{Z(0)} &=&\sum_{n=0}^{\infty}\left(\frac{-\beta
a_p}{2}\right)^n\frac{1}{n!}\prod_{i=1}^{n}\oint\frac{ds_i}{a}
\sum_{\left\{q_i\right\}}\left[\mbox{e}^{-i\sum_{i=0}^{n} q_i \varphi_i}
\left\langle \mbox{e}^{i\sum_{i=0}^n q_i\Theta_i}\right\rangle_0\right],
\end{eqnarray}
where we have defined charge $q_i=\pm p$ for each of the cosine terms in
Eq.~(\ref{pf2}) and denoted $\displaystyle{\sum_{\{q_i\}}}$ a sum over
all charge configurations. The thermal average can be evaluated exactly
using Wick's theorem;  we obtain the following equality:
\begin{eqnarray}
\left\langle \mbox{e}^{i\sum_{i=0}^n q_i\Theta_i}\right\rangle_0
=\mbox{e}^{-\frac{1}{2}\left\langle\left(\sum_{i=0}^n q_i\Theta_i\right)^2
\right\rangle_0}.\label{pf3}
\end{eqnarray}
The average $\langle\Theta^2\rangle_0$  encountered in Eq.~(\ref{pf3})
can be evaluated using the inverse of the microscopic length scale $1/a$
as the ultraviolet cutoff.  We have the following:
\begin{eqnarray}
\langle\Theta^2(\varphi)\rangle_0&=&\lim_{\varphi\rightarrow\varphi^\prime}
\langle\Theta(\varphi)\Theta(\varphi^\prime)\rangle_0.\nonumber\\
&=&\Delta\ln\frac{R_0}{a}.
\end{eqnarray}
The contributions of the charge configurations in which $\sum_iq_i\neq0$ to
the partition function are suppressed as a result of excess factors of
$\displaystyle{\exp\left( -\frac{p^2\Delta}{2}\ln\frac{R_0}{a}\right)}$.
Therefore, it is necessary to sum only over the configurations in which
there is
no net charge. Finally, the full partition function can be expressed as follows:
\begin{eqnarray}
\frac{Z(a_p)}{Z(0)} &=&\sum_{n=0}^{\infty}\left(\frac{-\beta
a_p}{2}\right)^{2n}\frac{1}{(2n)!}\prod_{i=1}^{2n}\oint\frac{ds_i}{a}
\sum_{\left\{\sum q_{i}=0\right\}}\mbox{e}^{-i\sum_{i=0}^{2n} q_i \varphi_i}
\mbox{e}^{-\frac{\Delta}{2}\sum_{i,j}q_iq_j\ln\frac{\vec{x}_i-\vec{x}_j}{a}},
\end{eqnarray}
\bottom{-2.7cm}
\begin{multicols}{2}
\hspace{-.15in}analogous to the expression of the partition function for a system of 2D
neutral coulomb gas.   The charge of the particles is $p\sqrt{\Delta/2}$ and
the particles are distributed along the circumference of the domain.
Following the Coulomb gas treatment for the 2D phase transitions\cite{Nien},
we derive flow equations for the running coupling constants,
\begin{eqnarray}
a\frac{d a_p(a)}{da}&=&\left[1-\frac{p^2\Delta(a)}{2}\right]a_p(a),\\
a\frac{d \Delta(a)}{da}&=&-2\pi^2\Delta^2(a)a_p^2(a).
\end{eqnarray}
For $a_p(a)\ll 1$, we have the following:
\begin{eqnarray}
\frac{a_p(a)}{a_p(a_0)}=\left(\frac{a}{a_0}\right)^{1-p^2\Delta/2}.
\end{eqnarray}
The scaling relation implies the relevancy, and the irrelevancy transition
temperature of $a_p(a)$ at $T_p$ is given by
\begin{eqnarray}
k_BT_p=\frac{4\pi\kappa}{p^2}.
\end{eqnarray}
As compared to the Kosterlitz-Thouless transition temperature
$k_BT_{KT}=\pi\kappa/2$,  $T_1$ and $T_2$ are above $T_{KT}$.
Our result is analogous to the scaling index of the symmetry-breaking
perturbation in the 2D planar model obtained by the spin-wave
approximation~\cite{RGvortices}.

In the low-temperature phase,  we consider fluctuations up to a
cutoff that is proportional to the sample size, or $a\sim R_0$,
and consider the renormalized coupling constant $a_p=a_p(R_0)R_0$.
We find the scaling relation,
\begin{eqnarray}
a_p&\sim&R_0^{-p^2\Delta/2}.
\end{eqnarray}
Based on a theory of fixed $\kappa$, the renormalized anisotropic line
tension decreases as a power law with $R_0$, the radius of the
boundary, with exponent $-p^2\Delta/2$.  Using this relation,  we
investigated the effect of thermal fluctuation on the domain boundary.  The
result is depicted in  plot of $\Psi$ at the maximum of $I$, $\Psi_0$, 
as a function of $R_0$ in Fig.~\ref{maxPsiT}.
We notice the rounding off at the maximum of $\Psi_0$  and the decrease in
the magnitude of $\Psi_0$ with increasing temperature.  The above
comparisons are made for domains with $\delta=0.4$ and $\gamma=0.38$
when $R_0=0.2$.

We can also look at the effect of fluctuations on $b$.  We let $a_n=0$
for all $n$ and the Hamiltonian becomes
\end{multicols}
\top{-2.8cm}
\begin{eqnarray}
H[\Theta]=\frac{\kappa}{2}\int\left\{|\nabla\Theta|^2+b
\left[\left(-\Theta_x^2+\Theta_y^2\right)\cos
2\Theta-2\Theta_x\Theta_y\sin 2\Theta\right]\right\}dA.
\end{eqnarray}
The partition function
\begin{eqnarray}
Z(b)=Z(0)\left\langle\exp\left[\frac{\beta
b\kappa}{4}\int\left(\Theta_{{\bar
z}}^2\mbox{e}^{i2\Theta}+\Theta_z^2\mbox{e}^{-i2\Theta}\right)dA\right]\right
\rangle_0,
\end{eqnarray}
\bottom{-2.7cm}
\begin{multicols}{2}
\hspace{-.15in}which leads to the flow equations similar to those for $a_k$,
\begin{eqnarray}
a\frac{d b(a)}{da}&=&-2\Delta(a)b(a),\\
a\frac{d \Delta(a)}{da}&=&-2\pi^2\Delta^2(a)b^2(a),
\end{eqnarray}
from which we deduce that $b$ is always irrelevant at finite temperature.

Following the same argument as for the case of $a_p$,  we consider fluctuations
that cut off at $a\sim R_0$ in the ordered phase,  and take the renormalized
coupling constant $b=b(R_0)$.  We find that the $b$ scales as $R_0^{-2\Delta}$ 
in a theory with fixed $\kappa$.  When thermal fluctuations are important, 
the magnitude $|b|$ is at its maximun ($\le1$) for the smallest 
\vadjust{\vskip 0.1in
\narrowtext
\begin{figure}
\centerline{\epsfig{file=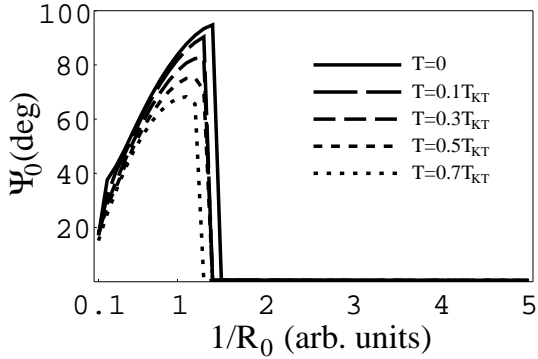,width=3in}}
\caption{Plots of $\Psi_0$  as a function of $R_0$ for
$\kappa=1$, $\sigma_0=4$ at $T=0$, $0.1T_{KT}$, $0.3T_{KT}$, $0.5T_{KT}$,
and $0.7T_{KT}$. At $R_0=0.2$,  all domains have the same 
renormalized coefficients in the expansion of the anisotropic line tension, 
namely $a_1=1.6$ and $a_2=0.6$. }
\label{maxPsiT}
\end{figure}
\vskip 0.05in}
$R_0$ and decreases as $R_0$ increases. Thermal effects reduce 
the already insignificant boundary correction due to $b$.  
This reinforces our earlier conclusion that $b$ has no significant 
influence on the domain boundary.

In summary, thermal fluctuations act to renormalize the anisotropic 
parameters.  The influence of thermal fluctuations on the boundary shape 
can be studied using the mean-field approximation with renormalized 
anisotropic parameters.  The boundary correction due to elastic 
anisotropy, which is negligible at $T=0$, further decreases as a result 
of thermal fluctuations.  The deviation of the boundary from a circle 
that results from line-tension anisotropy is displayed in terms of 
the domain-size dependence of the excluded angle in Fig.~\ref{andenf}. 
The maximum of the excluded angle,  which is higher in the 
experimental observations  than predicted at $T=0$ (see Fig.~\ref{anden}), 
will be reduced when thermal fluctuations are introduced.  This 
indicates that the line-tension anisotropy of the monolayer system under 
study may be very strong.

\section{Conclusions}
\label{sec:conclusions}
We have described in this paper a systematic investigation of a
system of a 2D ordered medium with a boundary.  Beginning with the free
energy Eq.~(\ref{sysenr}), which describes a bounded monolayer, we
have derived the Euler-Lagrange equations for the texture and the
boundary.  From the boundary conditions we have shown that the
boundary under consideration must be smooth.  A continuation of the
analysis in the spirit of Ref.  \cite{RudBru} for the cases of both
the domains and the bubbles reveals that bubbles do not remain
circular when the only term in the anisotropic line tension, as given
by Eq.~(\ref{expansion}), is $a_1\cos\phi$, while circular domains are
not affected.  There is, thus, no simple inversion symmetry between 
domains and bubbles, as one would intuitively expect.  Perturbative
calculations have been carried out to investigate the influence of the
$a_2$ term, the second-harmonic contribution in the line tension, and the
elastic anisotropy, parametrized by $b$, for the domains.
Assuming the domains are nearly circular and the anisotropies are
weak, the perturbed domain shapes are then computed analytically to 
first order in small parameters.  We have examined the boundary
response to the anisotropic parameters $b$ and $a_2$.  Our results for
the boundary response to $a_2$ are in qualitative agreement with those
reported in Refs.  \cite{GalaFour,numerics}.  We have also obtained the
dependence of the boundary shape on domain size that is in qualitative
agreement with experimental findings when $b=0$ \cite{FangTeer,cigar}.
As for the boundary response to $b$, our perturbative results contrast
with the conclusions arrived at in Ref.~\cite{GalaFour}.  Textural
correction plays an important role.  These conclusions are confirmed in
our numerical studies \cite{numerics,future}.  The $b$ contribution to
the boundary is much weaker than that of the line-tension
anisotropy.  The quality of the fit to the currently available
experimental observations is not sensitive to the value of $b$ in the
range from $-0.5$ to $0.5$.  
Considering only the line-tension anisotropy $a_2$, the result of
the perturbative analysis has
qualitatively captured the essential features of the experimental
observations. The quantitative mismatch can be attributed in part to the 
inapplicability of a perturbative approach to a parameter region in which 
the anisotropic parameters are large.  The detailed difference
between the simple model we adopt and the actual complex underlying
structure of the $L_2$ domains may also contribute to such
discrepancies.  Long-range dipolar repulsion has been ignored.
The tilt degree of freedom ~\cite{Tabe},  which may not be significant 
in the small domain regime but can be important in the large-$R_0$ region,
is not included. 

In the case of the bubbles, we evaluate the
boundary response due to the $a_1$ contribution in the line tension.
We are able to produce a dependence of the shape of the bounding curve
on bubble size that compares favorably with experimental
observations.  This result has been reported earlier in Ref.
\cite{FangTeer}.  
The parameters of the domains and those of the bubbles,  both obtained
with by-eye fit of the result of the analysis to the experimental data,
are in reasonable agreement,  taking into consideration the fact that
the data for the domains are taken in the $L_2/LE$ coexistence region and
those for the bubbles is obtained for gaseous bubbles surrounded by the 
$L_2$ phase. Finally, we have presented an analysis of the effect of
thermal fluctuations that leads to a domain-size dependence of the
line-tension and elastic anisotropies.  The analysis also suggests that 
the line-tension anisotropy may be very strong.  

\section*{Acknowledgments}

We are grateful to Professor Robijn Bruinsma, Dr. Jiyu Fang, and Ellis Teer 
for useful discussions.  We thank Professor H. Saleur and Professor P. 
Fendley for interesting ideas on the analysis of the effect of thermal 
fluctuations.  We also thank Professor R.B. Meyer for suggestions with 
regard to the depiction of textures.  We are especially indebted to
Professor Charles Knobler for enlightening discussions and for his
careful reading of the manuscript.

\appendix
\section{Geometry and coordinate systems}
\label{geom}
We enumerate in this section the forms taken by various geometrical
quantities of a 2D space curve in different coordinate systems and
 the relationships between those forms.  A curve $\Gamma$
surrounding $\Omega$ can be represented by a one-parameter trajectory
of the position vector $\vec{r}(t)$, where $t$ is parameter.  Its unit
tangent vector is given by $\hat{t}=d\vec{r}/ds$ and the unit outward
normal is given by $\hat{n}=\vec{n}/|\vec{n}|$, where $\vec{n}=\pm d^2
\vec{r}/ds^2$ and $ds\equiv |d\vec{r}/dt|dt$.  We let $\vartheta$ be the
angle between the normal vector of the curve and a reference axis.
Then the radius of curvature is $|ds/d\vartheta|$.

Consider the problem of a nearly circular domain.  The coordinate
system of choice is obviously plane-polar.  It is
convenient to use $\varphi$ as the independent variable.  We then
write $\Gamma$ as $\vec{r}(\varphi)=\mbox{e}^{k(\varphi)}\hat{\rho}$, where
$\mbox{e}^{k(\varphi)}$ is the radial distance from the origin and $\varphi$
is the polar angle at $\vec{r}$.  A length element given by
$ds=\mbox{e}^k\sqrt{1+k^{\prime2}}d\varphi$ is in the positive direction of
$\Omega$.  The unit vector $\vec{n}=-d^2\vec{r}/ds^2$ points away from
the origin, or outwards from $\Gamma$,
\begin{eqnarray}
\hat{t}=&&\frac{k^\prime\hat{\rho}+\hat{\varphi}}{\sqrt{1+k^{\prime2}}},\\
\hat{n}=&&\frac{\hat{\rho}-k^\prime\hat{\varphi}}{\sqrt{1+k^{\prime2}}}.
\end{eqnarray}
We have $\cos\vartheta=\hat{n}\cdot\hat{x}$, which gives
$\vartheta=\varphi-\tan^{-1}k^\prime$.  It follows that
\begin{eqnarray}
\frac{d\vartheta}{ds}=\left(1-\frac{k^{\prime\prime}}{1+k^{\prime2}}\right)\frac
{1}{\mbox{e}^k\sqrt{1+k^{\prime2}}}.
\end{eqnarray}
Similar relations can be derived for the case of a nearly circular
bubble.  Here, the length element in the positive direction of
$\Omega$ is $ds=-\mbox{e}^k\sqrt{1+k^{\prime2}}d\varphi$ and the outward
normal $\vec{n}= d^2\vec{r}/ds^2$.  We obtain
$\vartheta=\pi+\varphi-\tan^{-1}k^\prime$.  The geometrical quantities
are evaluated as follows:
\begin{eqnarray}
\hat{t}=&&-\frac{k^\prime\hat{\rho}+\hat{\varphi}}{\sqrt{1+k^{\prime2}}},\\
\hat{n}=&&-\frac{\hat{\rho}-k^\prime\hat{\varphi}}{\sqrt{1+k^{\prime2}}},\\
\frac{d\vartheta}{ds}=&&-\left(1-\frac{k^{\prime\prime}}{1+k^{\prime2}}
\right)\frac{1}{\mbox{e}^k\sqrt{1+k^{\prime2}}}.
\end{eqnarray}

Cartesian coordinates are useful when we are interested in a small
region of a large circular domain or bubble, on the scale of which
the boundary is nearly a straight line.  In Cartesian coordinates, we
have for the position vector $\vec{r}(t)\equiv x(t)\hat{x}
+y(t)\hat{y}$, where $\hat{x}$ and $\hat{y}$ are the unit basis
vectors.  We can always pick $y$ as the independent variable and
$x=\eta(y)$ as the dependent variable for the curve $\Gamma$.  The
symbol $\eta(y)$ is chosen deliberately to avoid confusion with the
independent variable $x$ for the texture $\Theta$.  We take $\Omega$
to reside in the region $x<0$, assume that $\Gamma$ nearly coincides
with the $y$ axis, and take the virtual defect to be in $x>0$ near the
$y$ axis as depicted in Fig.~\ref{ctst}.  We have
$ds=\sqrt{1+\eta^{\prime2}}dy$ traversing along the positive direction
of $\Gamma$ and $\vec{n}=- d^2\vec{r}/ds^2$.  We immediately obtain
\begin{eqnarray}
\hat{t}=&&\frac{\eta^\prime\hat{x}+\hat{y}}{\sqrt{1+\eta^{\prime2}}},\\
\hat{n}=&&\frac{\hat{x}-\eta^\prime\hat{y}}{\sqrt{1+\eta^{\prime2}}}.
\end{eqnarray}
The angle between the normal and the $x$ axis is
$\vartheta=-\tan^{-1}\eta^\prime$, and
\begin{eqnarray}
\frac{d\vartheta}{ds}=-\frac{\eta^{\prime\prime}}{\sqrt{1+\eta^{\prime
2}}^3}.
\end{eqnarray}

\section{The extremum equations}
We begin with the free energy Eq.~(\ref{sysenr}).  In Cartesian
coordinates, the elastic energy density ${\cal H}_b$ in Eq.
(\ref{enrdens}) is given as
\begin{eqnarray}
{\cal H}_b=\frac{\kappa}{2}\left\{|\nabla\Theta|^2 +
b\left[\left(-\Theta_x^2+\Theta_y^2\right)\cos
2\Theta-\right.\right.\nonumber\\
\left.\left.-\Theta_x\Theta_y\sin 2\Theta\right]\right\}.
\end{eqnarray}
Taking the variation of $H[\Theta]$ with respect to $\Theta$, we find
\begin{eqnarray}
\delta H=&&\int_\Omega\left[\frac{\partial {\cal H}_b}{\partial
\Theta}-\frac{\partial}{\partial x}\frac{\partial {\cal
H}_b}{\partial\Theta_x}-\frac{\partial}{\partial y}\frac{\partial
{\cal H}_b}{\partial\Theta_y}\right]\delta \Theta \;dA\pm\nonumber\\
&&\oint_\Gamma\left[-\frac{\partial {\cal
H}_b}{\partial\Theta_y}\frac{dx}{ds}+\frac{\partial {\cal
H}_b}{\partial\Theta_y}\frac{dy}{ds}\right]\delta\Theta\;ds-\nonumber\\
&&\oint_\Gamma
\sigma^\prime(\vartheta-\Theta)\delta\Theta\;ds.  \label{vart}
\end{eqnarray}
The boundary integral $\oint_\Gamma$ is taken counterclockwise.  The
$+$ in Eq.~(\ref{vart}) is appropriate to the case of domains while
$-$ is appropriate to bubbles.  The equilibrium condition $\delta
H/\delta \Theta=0$ results in the Euler-Lagrange equation
\begin{eqnarray}
\frac{\partial {\cal H}_b}{\partial \Theta}-\frac{\partial}{\partial
x}\frac{\partial {\cal
H}_b}{\partial\Theta_x}-\frac{\partial}{\partial y}\frac{\partial
{\cal H}_b}{\partial\Theta_y}=0
\end{eqnarray}
for $(x, y)\in\Omega$, which can be reduced to Eq.~(\ref{mbulk}).  The
boundary conditions can be expressed in Cartesian coordinates system
as follows:
\end{multicols}
\renewcommand{\thesection}{{\Alph{section}}}
\renewcommand{\theequation}{{\thesection.\arabic{equation}}}
%\top{-2.8cm}
\widetext
\begin{eqnarray}
\kappa
\left\{\Theta_x\frac{dy}{ds}-\Theta_y\frac{dx}{ds}+b\left[-\left(\Theta_y\frac{dx}{ds}+\Theta_x\frac{dy}{ds}\right)
\cos 2\Theta
%\right.\right.&&\nonumber\\
%\left.\left.
+\left(\Theta_x\frac{dx}{ds}-\Theta_y\frac{dy}{ds}\right)\sin 2
\Theta\right]\right\}\pm\sigma^\prime(\vartheta-\Theta)&&=0,
\end{eqnarray}

To display the derivation of the equilibrium equation for the bounding
curve $\Gamma$, it is more convenient to utilize
plane-polar coordinates.  In the case of a domain, we
rewrite the free energy in plane-polar coordinate as follows
\begin{eqnarray}
H[k]=\int_{-\pi}^{\pi}\left[\int_0^k{\cal H}_b
\mbox{e}^{2k_1}dk_1+\sigma(\varphi-\tan^{-1}k^\prime-\Theta)\mbox{e}^k\sqrt{1+k^
{\prime\,2}}
\right]d\varphi.
\end{eqnarray}
\bottom{-2.7cm}
\begin{multicols}{2}
\hspace{-.15in}We then take a variation of the free energy with respect to $k(\phi)$.
The equilibrium condition results in the Euler-Lagrange equation
\begin{eqnarray}
\frac{\partial{\cal H}}{\partial
k}-\frac{d}{d\varphi}\frac{\partial{\cal H}}{\partial
k^\prime}=0,\label{EulerGamma}
\end{eqnarray}
where
\begin{eqnarray}
{\cal H}[\varphi;k,k^\prime]&=&\int_0^k {\cal H}_b
\mbox{e}^{k_1}dk_1+\nonumber\\
&&\sigma(\varphi-\tan^{-1}k^\prime-\Theta)\mbox{e}^k\sqrt{1+k^{
\prime2}}.
\label{dHden}
\end{eqnarray}
To continue, we now look at the partial derivative of ${\cal H}$ with
respect to $k$. We have
\begin{eqnarray}
\frac{\partial {\cal H}}{\partial k}={\cal H}_b\mbox{e}^{2k}
+(\sigma-\sigma^\prime\Theta_k)\mbox{e}^k\sqrt{1+k^{\prime2}}.\label{fpart}
\end{eqnarray}
We have also the partial derivative of ${\cal H}$ with respect to
$k^\prime$,
\begin{eqnarray}
\frac{\partial {\cal H}}{\partial k^\prime}= \frac{(-\sigma^\prime
+\sigma k^\prime)\mbox{e}^k}{\sqrt{1+k^{\prime2}}}.  \label{dHdkp}
\end{eqnarray}
Taking the derivative of Eq.~(\ref{dHdkp}) with respect to the independent
variable $\varphi$ results in
\end{multicols}
\top{-2.8cm}
\begin{eqnarray}
\frac{d}{d\varphi}\frac{\partial {\cal H}}{\partial
k^\prime}=&&\left[\frac{\sigma^{\prime\prime}\left(\Theta_kk^\prime+
\Theta_\varphi\right)-\sigma^\prime
k^\prime\left(\Theta_kk^\prime+\Theta_\varphi\right)}{\sqrt{1+k^{\prime2}}}-
\frac{\sigma
k^{\prime2}+\sigma^{\prime\prime}}{\sqrt{1+k^{\prime2}}}-\frac{(\sigma+
\sigma^{\prime\prime})k^{\prime\prime}}{\sqrt{1+k^{\prime2}}^3}\right]
\mbox{e}^k.
\label{spart}
\end{eqnarray}
\bottom{-2.7cm}
\begin{multicols}{2}
\hspace{-.15in}
Equation (\ref{polar}) for the case of a domain follows immediately from
the substitution of Eqs.  (\ref{fpart}) and (\ref{spart}) into the
Euler-Lagrange equation, Eq.~(\ref{EulerGamma}).  We have assumed
here that the curve joins smoothly at $\varphi=-\pi, \pi$ and there is
no boundary contribution from these two end points.  However, we are
particularly interested in finding out if a cusp, in the form of a
discontinuity in the slope of the bounding curve, exists.  In our
system, which is symmetric about the $x$ axis, the singularity is 
expected to
occur on the $x$ axis.  We thus allow for the possibility that
$\Gamma$ has a discontinuity
in slope at $\varphi=0$ and determine the condition for the
discontinuity.  The assumption of a discontinuity in $\Gamma$ gives
rise to an extra boundary condition at $\varphi=0$,
\begin{eqnarray}
\left.\frac{\partial {\cal H}}{\partial k^\prime}\right|_{0^+}=0.
\end{eqnarray}
Using Eq.~(\ref{dHdkp}) and the fact that $\Theta(\mbox{e}^k)=0$, we thus get
\begin{eqnarray}
k^\prime=\frac{\sigma^\prime(-\tan^{-1}k^\prime)}{\sigma(-\tan^{-1}k^\prime)}.
\label{cusp}
\end{eqnarray}
For the line tension given in the form of Eq.~(\ref{expansion}),
$k^\prime=0$ is always a solution to Eq.~(\ref{cusp}).  In order for
$k^\prime$ to be nonzero at $\varphi=0$, it is necessary that the
slope of the right-hand side as a function of $k^\prime$ at
$k^\prime=0$ be greater that unity, i.e.,
\begin{eqnarray}
\left.\frac{d}{dk^\prime}\frac{\sigma^\prime(-\tan^{-1}k^\prime)}
{\sigma(-\tan^{-1}k^\prime)}\right|_{k^\prime=0}\geq1.
\end{eqnarray}
This implies $\sigma(0)\,[\sigma(0)+\sigma^{\prime\prime}(0)] \leq 0$.
We have excluded such a condition, in that it requires either
$|a_1|\ge\sigma_0$ or $3a_2\ge\sigma_0$, both of which are well beyond
the parameter regime that we are focusing on.  The boundary $\Gamma$
that we are solving for will not have a singularity.

In the case of a bubble,  we have
\begin{eqnarray}
{\cal H}[\varphi;k,k^\prime]&=&\int_k^\infty {\cal H}_b
\mbox{e}^{k_1}dk_1+\nonumber\\
&&\sigma(\pi+\varphi-\tan^{-1}k^\prime-\Theta)\mbox{e}^k\sqrt{1
+k^{\prime2}}.
\end{eqnarray}
This is similar to Eq.~(\ref{dHden}) apart from the limit of integration
in the first term on the right-hand side of the equation.

In the case of the Cartesian coordinate system, we write the free energy as
\begin{eqnarray}
H[\eta]&=&\int_{-\infty}^{\infty}\left[\int_{-\infty}^\eta{\cal H}_b
dx+\sigma(\vartheta-\Theta)\sqrt{1+\eta^{\prime\,2}}\right]dy
\end{eqnarray}
and the equilibrium condition can be obtained using the following
Euler-Lagrange equation:
\begin{eqnarray}
\frac{\partial {\cal H}}{\partial\eta}-\frac{d}{dy}\frac{\partial
{\cal H}}{\partial \eta^\prime}=0,
\end{eqnarray}
where
\begin{eqnarray}
{\cal H}(\varphi;\eta,\eta^\prime)=\int_{-\infty}^\eta{\cal H}_b
dx+\sigma(\vartheta-\Theta)\sqrt{1+\eta^{\prime\,2}}.
\end{eqnarray}

\end{multicols}
\top{-2.8cm}

\section{Sample Calculation}
\label{samplecalc}
We present here an analysis of the equilibrium equations for the special case
in which $b=0$ and only a single $a_p>0$.  In this case, the bulk
equation is automatically satisfied if we write $\Theta$ in the form
of Eq.~(\ref{harmonic}).  We first consider the case of a domain,
for which the boundary condition is given by Eq.~(\ref{boundd}).  We
substitute the boojum texture $\Theta_0$ and find
%\top{-2.8cm}
\begin{eqnarray}
&&\frac{\kappa\alpha
\mbox{e}^{pk}}{iR_0}\left[-\frac{\mbox{e}^{ip\varphi}}{1-\alpha
\mbox{e}^{p(k+i\varphi)}}+\frac{ \mbox{e}^{-ip\varphi}}{1-\alpha
\mbox{e}^{p(k-i\varphi)}}\right]-\frac{a_p
i}{2}\left[\frac{\mbox{e}^{ip\varphi}-\alpha \mbox{e}^{pk}}{1-\alpha
\mbox{e}^{p(k+i\varphi)}}-\frac{\mbox{e}^{-ip\varphi}-\alpha
\mbox{e}^{pk}}{1-\alpha
\mbox{e}^{p(k-i\varphi)}}\right]=0,\nonumber\\
\Rightarrow&&\frac{\kappa}{iR_0}\left[-\frac{1}{1-\alpha
\mbox{e}^{p(k+i\varphi)}}+\frac{1}{1-\alpha \mbox{e}^{p(k-i\varphi)}}\right]-
\frac{a_p i}{2\alpha \mbox{e}^{pk}}\left[\frac{1-\alpha^2
\mbox{e}^{2pk}}{1-\alpha
\mbox{e}^{p(k+i\varphi)}}-\frac{1-\alpha^2 \mbox{e}^{2pk}}{1-\alpha
\mbox{e}^{p(k-i\varphi)}}\right]=0,\nonumber\\
\Rightarrow &&(\alpha^2 \mbox{e}^{2pk}+2\epsilon\alpha \mbox{e}^{pa}
-1)\left[\frac{1}{1-\alpha \mbox{e}^{p(k+i\varphi)}}-\frac{1}{1-\alpha
\mbox{e}^{p(k-i\varphi)}}\right]=0.  \label{redb}
\end{eqnarray}
The above boundary condition is to be satisfied for all $(e ^k, \varphi)\in
\Gamma$ and hence the coefficient in Eq.~(\ref{redb}) has to be 0,
which gives $\alpha \mbox{e}^{pk}=-\epsilon \pm\sqrt{1+\epsilon^2}$,
where $\epsilon\equiv \kappa/(p a_p R_0)$. Together
with the requirement that $\Theta$ does not have a singularity in
$\Omega$, we arrive at Eq.~(\ref{defpos}), where $R_0=\mbox{e}^k$ is the
radius of the circular boundary.  As for the case where $a_p<0$,
we substitute $\Theta_-=\Theta_0+\pi/p$ into the boundary condition
Eq.~(\ref{boundd}).  Requiring that the texture is continuous in
$\Omega$, we find Eq.~(\ref{defpos}) with
$\epsilon\equiv\kappa/(p |a_p|R_0)$.  The results for bubbles can be
obtained in a similar manner.  Although we have demonstrated a
solution of $\Theta$ for $p$ any integer, a circular domain shape
does not satisfy the equilibrium condition for $\Gamma$ in general.

We now examine the domain shape for the case in which $p=1$ by
substituting the boojum texture and a circular boundary into Eq.
(\ref{polar}).  We have
\begin{eqnarray}
\frac{\kappa}{2}|\nabla\Theta_0|^2&=&\frac{\kappa}{2}
\left\{-\left[\mbox{e}^{i\varphi}f_0^\prime(\mbox{e}^{k+i\varphi})-\mbox{e}^{-i
\varphi}f_0^\prime(\mbox{e}^{
k-i\varphi})\right]+\left[\mbox{e}^{i\varphi}f_0^\prime(\mbox{e}^{k+i\varphi})+
\mbox{e}^{-i\varphi}f_
0^\prime(\mbox{e}^{k-i\varphi})\right]\right\}\nonumber\\
&=&2\kappa
f_0^\prime(\mbox{e}^{k+i\varphi})f_0^\prime(\mbox{e}^{k-i\varphi})\nonumber\\
&=&\frac{2\kappa\alpha^2}{(1-\alpha \mbox{e}^{k+i\varphi})(1-\alpha
\mbox{e}^{k-i\varphi})}\nonumber\\
&=&\frac{2\kappa\alpha^2}{1-\alpha^2\mbox{e}^{2k}}\left(\frac{\alpha
\mbox{e}^{k+i\varphi}}{1-\alpha \mbox{e}^{k+i\varphi}}+\frac{\alpha
\mbox{e}^{k-i\varphi}}{1-\alpha \mbox{e}^{k-i\varphi}}+1\right)\nonumber\\
&=&a_1\alpha\left(\frac{1}{1-\alpha \mbox{e}^{k+i\varphi}}+\frac{1}{1-\alpha
\mbox{e}^{k-i\varphi}}-1\right)\label{Gamma1}
\end{eqnarray}
and
\begin{eqnarray}
&&\sigma^\prime(\varphi-\Theta_0)\Theta_{0k}+\sigma^{\prime\prime}
(\varphi-\Theta_0)\Theta_{0\varphi}\nonumber\\
=&&\frac{a_1}{2}\left\{\left[\mbox{e}^{i(\varphi-\Theta_0)}-\mbox{e}^{-i(\varphi
-\Theta_0)}
\right]\left[\mbox{e}^{k+i\varphi}f_0^\prime(\mbox{e}^{k+i\varphi})-\mbox{e}^{k-
i\varphi}f_0^{\prime}
(\mbox{e}^{k-i\varphi})\right]\right.\nonumber\\
&&\left.-\left[\mbox{e}^{i(\varphi-\Theta_0)}+\mbox{e}^{-i(\varphi-\Theta_0)}
\right]
\left[\mbox{e}^{k+i\varphi}f_0^\prime(\mbox{e}^{k+i\varphi})+\mbox{e}^{k-i
\varphi}f_0^{\prime}
(\mbox{e}^{k-i\varphi})\right]\right\}\nonumber\\
=&&-a_1\mbox{e}^k\left[\mbox{e}^{i\Theta}f_0^\prime(\mbox{e}^{k+i\varphi})+\mbox
{e}^{-i\Theta}f_0^{\prime}
(\mbox{e}^{k-i\varphi})\right]\nonumber\\
=&&a_1\alpha \mbox{e}^k\left(\frac{1}{1-\alpha
\mbox{e}^{k+i\varphi}}+\frac{1}{1-\alpha
\mbox{e}^{k-i\varphi}}\right)\label{Gamma2}
\end{eqnarray} 
%\bottom{-2.7cm}
\begin{multicols}{2}
\hspace{-.15in}Substituting Eqs.  (\ref{Gamma1}) and (\ref{Gamma2}) into Eq.
(\ref{polar}), all the $\varphi$ dependence cancels exactly.  Equality
is achieved by picking $\lambda=a_1 \alpha \mbox{e}^k-a_0$.  We conclude that
the circular domain with boojum texture is an equilibrium configuration for
the case where $p=1$.  It is obvious that this is in general not true
for any other $p\neq1$.

We now examine the case of the bubble for $p=1$.  We substitute the
inverse boojum $\Theta_i$ into Eq.~(\ref{polar}).
We compute the following when $\hat{c}$ directors on the boundary favor
pointing
in the bulk, or pointing away from the center of the bubble, or $a_1>0$,
\begin{eqnarray}
&&\sigma^\prime(\pi+\varphi-\Theta_i)\Theta_{ik}+\sigma^{\prime\prime}(\pi+
\varphi-\Theta_i)\Theta_{i\varphi}\nonumber\\
=&&\frac{a_1\alpha
}{\mbox{e}^k}\left[\frac{\mbox{e}^{i\varphi}(\mbox{e}^{i\varphi}-\alpha
\mbox{e}^{-k})}{(1-\alpha
\mbox{e}^{-k+i\varphi})^2}+\frac{\mbox{e}^{-i\varphi}(\mbox{e}^{-i\varphi}-
\alpha\mbox{e}^{-k})}{(1-\alpha \mbox{e}^{-k-i\varphi})^2}\right].
\end{eqnarray}
When the $\hat{c}$ directors on the boundary favor pointing away from the
bulk, or pointing toward the center of the bubble, or $a_1<0$, we
obtain
\begin{eqnarray}
&&\sigma^\prime(\varphi-\Theta_i)\Theta_{ik}+\sigma^{\prime\prime}
(\varphi-\Theta_i)\Theta_{i\varphi}\nonumber\\
=&&\frac{|a_1|\alpha
}{\mbox{e}^k}\left[\frac{\mbox{e}^{i\varphi}(\mbox{e}^{i\varphi}-\alpha
\mbox{e}^{-k})}{(1-\alpha
\mbox{e}^{-k+i\varphi})^2}+\frac{\mbox{e}^{-i\varphi}(\mbox{e}^{-i\varphi}-
\alpha\mbox{e}^{-k})}{(1-\alpha \mbox{e}^{-k-i\varphi})^2}\right].
\end{eqnarray}
And the following contribution is independent of the sign of $a_1$
\begin{eqnarray}
\frac{\kappa}{2}|\nabla\Theta_i|^2&=&\frac{|a_1|\alpha}{\mbox{e}^{2k}}\left(\frac{
1}{1-\alpha\mbox{e}^{-k+i\varphi}}+\right.\nonumber\\
&&\left.\frac{1}{1-\alpha \mbox{e}^{-k-i\varphi}}-1\right).
\end{eqnarray}
Approximating $\Theta_n\approx-\Theta_k$, $\Theta_t\approx-\Theta_\varphi$,
and putting together the above contributions, we find
\begin{eqnarray}
k^{\prime\prime}&=&-\frac{a_1\alpha}{\sigma_0\mbox{e}^k}\left[\frac{1}{1-\alpha
\mbox{e}^{-k+i\varphi}}+\frac{\mbox{e}^{i\varphi}(\mbox{e}^{i\varphi}-\alpha 
\mbox{e}^{-k})}
{(1-\alpha \mbox{e}^{-k+i\varphi})^2}+\mbox{c.c.}\right]\nonumber\\
&=&-\frac{a_1\alpha}{\sigma_0\mbox{e}^k}\left[\frac{\mbox{e}^{i2\varphi}-2
\alpha\mbox{e}^{-k+i\varphi}+1}
{(1-\alpha \mbox{e}^{-k+i\varphi})^2}+\mbox{c.c.}\right]\nonumber\\
&=&-\frac{a_1\alpha}{\sigma_0\mbox{e}^k}\left[\frac{(1-\alpha^2\mbox{e}^{-2k})
\mbox{e}^{i2\varphi}}
{(1-\alpha \mbox{e}^{-k+i\varphi})^2} + 1 +\mbox{c.c.}\right]\nonumber\\
&=&-\frac{2a_1\alpha^2\epsilon}{\sigma_0\mbox{e}^{2k}}\left[\frac{\mbox{e}^{i2
\varphi}}{(1-\alpha \mbox{e}^{-k+i\varphi})^2} + 1 +\mbox{c.c.}\right],
\end{eqnarray}
which reduces to Eq.~(\ref{kbubble}).  We keep in the expressions of
$k^{\prime\prime}$ only the apparently nontrivial terms and drop
arbitrarily the constant terms for convenience.  In the actual
evaluation of $k$,   the constant is reinserted into
$k^{\prime\prime}$ to enforce the symmetry requirement
$k^\prime(\varphi)=-k^\prime(-\varphi)$.  The boundary correction
is taken to have little modification to the overall area
enclosed and the treatment for fixing the area using the
Lagrange multiplier $\lambda$ is ignored.

\section{Evaluation of the thermal averages}
In this appendix,  we describe the detail evaluation of
$\langle\Theta(\varphi)\Theta(\varphi^\prime)\rangle_0$
and $\langle\exp[i\sum q_i\Theta(\varphi_i)]\rangle_0$
that we encountered in Sec.~\ref{sec:thermal}. The
average $\langle{\cal O}\rangle_0$ is taken with respect to
\begin{eqnarray}
H_0[\Theta]=\frac{\kappa}{2}\int_\Omega dA |\nabla\Theta|^2.
\end{eqnarray}
We first note that the extremum equation for $\Theta$ is
given by Laplace's equation $\nabla^2\Theta=0$ in which
the solution accommodates any boundary condition.
Hence we can write $\Theta(x, y)=\Theta_1(x, y)+\Theta_2(x, y)$
in general, where
\begin{eqnarray}
\nabla^2\Theta_1&=&0,\label{extremal}\\
\left.\Theta_2\right|_\Gamma&=&0.
\end{eqnarray}
It can be shown further that $\Theta_1$ indeed minimizes $H_0[\Theta]$.
At low temperature, contributions from large $\Theta_2$ are suppressed
in the partition function.  We shall assume $\Theta_2\ll\Theta_1$ and
$H_0[\Theta]$ is second order in $\Theta_2$.  The quantity $\Theta_2$
can be neglected and we have, in the case of a domain,
\begin{eqnarray}
\Theta=\sum_{m=1}^\infty\rho^m(a_m
\mbox{e}^{im\varphi}+a_m^*\mbox{e}^{-im\varphi}).
\end{eqnarray}
$H_0[\Theta]$ can then be evaluated,
\begin{eqnarray}
H_0=4\pi\sum_{m=1}^\infty m|A_m|^2,
\end{eqnarray}
where $A_m=a_mR_0$.

We let $\Theta_0(\varphi)=\Theta(R_0\cos\varphi, R_0\sin\varphi)$,
\end{multicols}
\top{-2.8cm}
\begin{eqnarray}
&&\langle\Theta_0(\varphi)\Theta_0(\varphi^\prime)\rangle_0\nonumber\\
&=&\frac{\int({\cal D}\Theta)\Theta_0(\varphi)\Theta_0(\varphi^\prime)
\mbox{e}^{-\beta H_0}}{\int({\cal D}\Theta)\mbox{e}^{-\beta H_0}}\nonumber\\
&=&\frac{\prod_n\int dA_n dA_n^*\sum_k\sum_{k^\prime}(A_k\mbox{e}^{ik\varphi}+
A_k^*\mbox{e}^{-ik\varphi})(A_{k^\prime}\mbox{e}^{i{k^\prime}{\varphi^\prime}}
+A_{k^\prime}^*\mbox{e}^{-i{k^\prime}{\varphi^\prime}})
\mbox{e}^{-2\pi\beta\kappa\sum_mA_mA_m^*}}
{2\prod_n\int dA_n dA_n^*\mbox{e}^{-2\pi\beta\kappa\sum_mA_mA_m^*}}\nonumber\\
&=&\sum_{k=1}\left[\mbox{e}^{ik(\varphi-\varphi^\prime)}+\mbox{e}^{-ik(\varphi-
\varphi
^\prime)}\right]\frac{\int
dA_kdA_k^*A_kA_k^*\mbox{e}^{-2\pi\beta\kappa kA_kA_k^*}}{2\int
dA_kdA_k^*\mbox{e}^{-2\pi\beta\kappa kA_kA_k^*}}\nonumber\\
&=&\frac{\Delta}{2}\sum_{k=1}\frac{\mbox{e}^{ik(\varphi-\varphi^\prime)}+
\mbox{e}^{-ik(\varphi-\varphi^\prime)}}{k}\nonumber\\
&=&-\frac{\Delta}{2}\left\{\ln\left[1-\mbox{e}^{i(\varphi-\varphi^\prime)}\right
]
+\ln\left[1-\mbox{e}^{-i(\varphi-\varphi^\prime)}\right]\right\}\nonumber\\
&=&-\frac{\Delta}{2}\ln\left|4\sin^2\frac{\varphi-\varphi^\prime}{2}\right|
\nonumber\\
&=&-\Delta\ln\frac{|\vec{x}-\vec{x}^\prime|}{R_0}.
\end{eqnarray}
When $\varphi\rightarrow\varphi^\prime$, we then introduce an ultraviolet
cutoff
$R_0/a$ in the sum
\begin{eqnarray}
\langle\Theta^2(\varphi)\rangle_0=\Delta\sum_{k=1}^{R_0/a}\frac{1}{k}
=\Delta\ln\frac{R_0}{a}.
\end{eqnarray}

Evaluation of $\langle\exp[i\sum_i q_i \Theta(\varphi_i)]\rangle_0$
is best illustrated with
\begin{eqnarray}
&&\langle\mbox{e}^{i[\Theta(\varphi_1)-\Theta(\varphi_2)]}\rangle_0\nonumber\\
&=&\sum_{m_1=0}\sum_{m_2=0}\frac{1}{m_1!m_2!}\langle[i\Theta(\varphi_1)]^{m_1}
[-i\Theta(\varphi_2)]^{m_2}\rangle_0.\label{exponent}
\end{eqnarray}
We will apply Wick's theorem to compute the above average.  We first note
that the average vanishes when  $m_1+m_2$ is odd.  When both $m_1$ and
$m_2$ are even,  we have
\begin{eqnarray}
&&\langle[i\Theta(\varphi_1)]^{m_1}[-i\Theta(\varphi_2)]^{m_2}\rangle_0\nonumber
\\
&=&m_1!m_2!\sum_{l\mbox{ even}}
\frac{\langle \Theta(\varphi_1)\Theta(\varphi_2)\rangle_0^l}{l!}
\frac{\langle-\Theta^2(\varphi_1)\rangle_0^{\frac{m_1-l}{2}}}
{(\frac{m_1-l}{2})!2^{\frac{m_1-l}{2}}}
\frac{\langle-\Theta^2(\varphi_2)\rangle_0^{\frac{m_2-l}{2}}}
{(\frac{m_2-l}{2})!2^{\frac{m_2-l}{2}}}\nonumber\\
&=&(2n_1)!(2n_2)!\sum_{n_0=0}
\frac{\langle \Theta(\varphi_1)\Theta(\varphi_2)\rangle_0^{2n_0}}{(2n_0)!}
\frac{\langle-\Theta^2(\varphi_1)\rangle_0^{n_1-n_0}}
{(n_1-n_0)!2^{n_1-n_0}}
\frac{\langle-\Theta^2(\varphi_2)\rangle_0^{n_2-n_0}}
{(n_2-n_0)!2^{n_2-n_0}},
\end{eqnarray}
where $m_1=2n_1$, $m_2=2n_2$, and $l=2n_0$.   Similarly we can obtained for
the cases where both $m_1$ and $m_2$ are odd as below,
\begin{eqnarray}
&&\langle[i\Theta(\varphi_1)]^{m_1}[-i\Theta(\varphi_2)]^{m_2}\rangle_0\nonumber
\\
&=&(2n_1+1)!(2n_2+1)!\sum_{n_0=0}
\frac{\langle \Theta(\varphi_1)\Theta(\varphi_2)\rangle_0^{2n_0+1}}{(2n_0+1)!}
\frac{\langle-\Theta^2(\varphi_1)\rangle_0^{n_1-n_0}}
{(n_1-n_0)!2^{n_1-n_0}}
\frac{\langle-\Theta^2(\varphi_2)\rangle_0^{n_2-n_0}}
{(n_2-n_0)!2^{n_2-n_0}}
\end{eqnarray}
where $m_1=2n_1+1$, $m_2=2n_2+1$, and $l=2n_0+1$.  Combining these contributions
in Eq.~(\ref{exponent}),   we get
\begin{eqnarray}
&&\langle\mbox{e}^{i[\Theta(\varphi_1)-\Theta(\varphi_2)]}\rangle_0\nonumber\\
&=&\sum_{n_0=0}\sum_{n_1=0}\sum_{n_2=0}
\frac{\langle \Theta(\varphi_1)\Theta(\varphi_2)\rangle_0^{n_0}}{n_0!}
\frac{\langle -\Theta^2(\varphi_1)\rangle_0^{n_1}}
{n_1!2^{n_1}}
\frac{\langle -\Theta^2(\varphi_2)\rangle_0^{n_2}}
{n_2!2^{n_2}}\nonumber\\
&=&\mbox{e}^{\langle\Theta(\varphi_1)\Theta(\varphi_2)\rangle_0}
\mbox{e}^{-\frac{\langle\Theta(\varphi_1)\Theta(\varphi_2)\rangle_0}{2}}
\mbox{e}^{-\frac{\langle\Theta(\varphi_1)\Theta(\varphi_2)\rangle_0}{2}}
\nonumber\\
&=&\mbox{e}^{-\frac{[\Theta(\varphi_1)-\Theta(\varphi_2)]^2}{2}}.
\end{eqnarray}
\bottom{-2.7cm}
\begin{multicols}{2}
\hspace{-.15in}Although more  tedious enumerations of the combinations of the correlation
functions must be carried out in order to evaluate  $\langle\exp[i\sum_i q_i
\Theta(\varphi_i)]\rangle_0$, the same principle applies.

It can be observed in this simple case that the contribution in
the partition function Eq.~(\ref{partitionfunction}) of
\begin{eqnarray}
&&\langle\mbox{e}^{i[\Theta(\varphi_1)+\Theta(\varphi_2)]}\rangle_0\nonumber\\
&=&\mbox{e}^{-\langle\Theta(\varphi_1)\Theta(\varphi_2)\rangle_0}
\mbox{e}^{-\frac{\langle\Theta(\varphi_1)\Theta(\varphi_2)\rangle_0}{2}}
\mbox{e}^{-\frac{\langle\Theta(\varphi_1)\Theta(\varphi_2)\rangle_0}{2}}
\nonumber\\
&=&\mbox{e}^{-2\Delta\ln\frac{R_0}{a}}
\mbox{e}^{-\Delta\ln\frac{\vec{x}-\vec{x}^\prime}{a}}
\end{eqnarray}
is approximately a factor $\mbox{e}^{-2\Delta\ln\frac{R_0}{a}}$ smaller
than that of
\begin{eqnarray}
\langle\mbox{e}^{i[\Theta(\varphi_1)-\Theta(\varphi_2)]}\rangle_0=\mbox{e}^{
\Delta\ln\frac{\vec{x}-\vec{x}^\prime}{a}}.
\end{eqnarray}

\end{multicols}

\begin{references}
\bibitem{phases} C.M. Knobler and R.C. Desai, Annu. Rev. Phys. Chem.
{\bf 43}, 207 (1992).

\bibitem{Mermin} N.D. Mermin, in {\em Quantum Fluids and Solids},
edited by S.B. Trickey, E. Adams, and J. Duffy (Plenum, New York,
1977).

\bibitem{LanSeth} S.A. Langer and J.P. Sethna, Phys.  Rev.  A
{\bf 34}, 5035 (1986).

\bibitem{RivMeu} S. Rivi\`{e}re and J.
Meunier, Phys.  Rev.  Lett.  {\bf 74}, 2495 (1995).

\bibitem{FangTeer} J. Fang, E. Teer, C.M. Knobler, K.-K. Loh, and J.
Rudnick, Phys.  Rev.  E {\bf 56}, 1859 (1997).

\bibitem{cardioid} G.
Brezesinski, E. Scalas, B. Struth, H. M\"{o}hwald, F. Bringezu, U.
Gehlert, G. Weidemann, and D. Vollhardt, J. Phys.  Chem.  {\bf 99},
8758 (1995).

\bibitem{cigar} J. Fang and C.M. Knobler (unpublished).

\bibitem{FiscBru} T.M. Fischer, R.F. Bruinsma, and C.M. Knobler, Phys.
Rev.  E {\bf 50}, 413 (1994).

\bibitem{RudBru} J. Rudnick and R.
Bruinsma, Phys.  Rev.  Lett.  {\bf 74}, 2491 (1995).

\bibitem{GalaFour} P. Galatola and J.B. Fournier, Phys.  Rev.
Lett.  {\bf 75}, 3297 (1995).

\bibitem{Saleur}  P. Fendley and H. Saleur, Phys. Rev. Lett.
{\bf 75}, 4492 (1995).

\bibitem{private} J. Rudnick and K.-K. Loh  (private communications).

\bibitem{numerics} K.-K. Loh and J. Rudnick, Phys. Rev. Lett. {\bf 81},
4935 (1998).

\bibitem{Teer} E. Teer and C.M. Knobler (private communications);
C. Lautz, T.M. Fischer, M. Weygand, M. L\"{o}sche, P.B. Howes, and K.
Kjaer, J. Chem. Phys. {\bf 108}, 4640 (1998).

\bibitem{defects} P. Chaikin and T. Lubensky, {\em Priciples of Condensed
Matter
Physics} (Cambridge University Press, Cambridge, 1995).

\bibitem{PettyLuben} D. Pettey and T.C. Lubensky, Phys. Rev. E
{\bf 59}, 1834 (1999).

\bibitem{GradRyzh} I.S. Gradshteyn and I.W. Ryzhik, {\em Table of Integrals,
Series and Products} (Academic Press, London, 1965).

\bibitem{future}
K.-K. Loh and J. Rudnick (unpublished).

\bibitem{Tabe} Y. Tabe, N. Shen, E. Mazur, and H. Yokoyama, Phys.
Rev. Lett. {\bf 82}, 759 (1999).

\bibitem{HP} We are grateful to Professors H. Saleur and P.
Fendley for generously providing us with this idea for the analysis of
thermal fluctuations.

\bibitem{Nien} B. Nienhuis, in {\em Phase Transitions and Critical
Phenomena}, Vol.  11, edited by C. Domb and J. Lebowitz (Academic
Press, London, 1987).

\bibitem{RGvortices} J.V. Jos\'{e}, L.P. Kadanoff, S. Kirkpatrick and D.R.
Nelson, Phys. Rev. B {\bf 16}, 1217 (1977).

\end{references}
\end{document}